\documentclass[preprint,floatfix] {revtex4}

\newcommand{\Hvec}{\mathrm {\mathbf {H}}}

\newcommand{\rvec}{\mathrm {\mathbf {r}}} 

\newcommand{\lvec}{\mathrm {\mathbf {l}}}
 
\usepackage{graphicx}
\usepackage{subfigure}
\usepackage{xcolor}
\usepackage{amsmath}
\usepackage{enumerate}
\usepackage{longtable}
\usepackage{tabularx}
\usepackage{multirow}
\usepackage{amssymb}
\usepackage{color, soul}
\usepackage[utf8]{inputenc}
\usepackage{amsfonts}
\usepackage{hyperref}
\definecolor{darkblue}{rgb}{0,0,0.5}
\setulcolor{darkblue}

\begin{document}

\title{{Density functional study of atoms spatially confined inside a hard sphere}}

\author{Sangita Majumdar}
\author{Amlan K.~Roy}
\altaffiliation{Corresponding author. Email: akroy@iiserkol.ac.in, akroy6k@gmail.com.}
\affiliation{Department of Chemical Sciences, Indian Institute of Science Education and Research \\
Kolkata, Mohanpur-741246, Nadia, WB, India}


\begin{abstract}
{An atom placed inside a cavity of finite dimension offers many interesting features, and thus has been a topic of great 
current activity. This work proposes a density functional approach to pursue both ground and excited states of a 
multi-electron atom under a spherically impenetrable enclosure. The radial Kohn-Sham (KS) equation has been solved by 
invoking a physically motivated work-function-based exchange potential, which offers near-Hartree-Fock-quality results. 
Accurate numerical eigenfunctions and eigenvalues are obtained through a generalized pseudospectral method (GPS) fulfilling 
the Dirichlet boundary condition. Two correlation functionals, \emph{viz.,} (i) simple, parametrized local 
Wigner-type, and (ii) gradient- and Laplacian-dependent non-local Lee-Yang-Parr (LYP) functionals are adopted 
to analyze the electron correlation effects. Preliminary exploratory results are offered for ground states of 
He-isoelectronic series ($Z=2-4$), as well as Li and Be atom. Several low-lying 
singly excited states of He atom are also reported. These are compared with available literature results--which 
offers excellent agreement. Radial densities as well as expectation values are also provided. 
The performance of correlation energy functionals are discussed critically. In essence, this presents a simple, 
accurate scheme for studying atomic systems inside a \emph{hard} spherical box within the rubric of KS density 
functional theory.}



\end{abstract}
\maketitle

\section{Introduction} 
In recent years, we have witnessed a proliferation of research activity in the field of quantum confinement. 
A particularly interesting situation arises when a spatial barrier causes dramatic changes in observable 
properties compared to their free counterpart. This type of perturbation deeply influences the ionization threshold, 
atomic size, molecular bond size, polarizability, energy spectrum etc. This remarkable difference in observed 
physico-chemical properties of two systems have motivated considerable amount of theoretical and experimental 
works in this direction. They have found significant relevance in modeling a large range of physical and 
chemical systems, \emph{viz.}, calculation of zero-point energy in fluids at high density, description of 
magnetic behavior of metals under small magnetic field, restricted rotator, so-called \emph{artificial atom} 
or quantum dot, matter (atom, molecule, ion), trapped atoms/molecules (inside cavity, zeolite channel, or 
in endohedral fullerene cage), etc. Particle in a box with impenetrable walls appears in some applications 
of acoustics and biology as well. Apart from the academic appeal, there is an underlying technological bearing 
in connection to the development of new materials with unconventional properties. The field is quite vast, 
relatively young and continuously expanding. Many excellent, elegant reviews are available. The interested 
reader may consult following works and references therein \cite{jaskolski96, sabin09, sen14, leykoo18}. 

Different models were proposed in literature to probe the electronic structure of confined atoms. A simple 
intuitive one to represent this is to place the atom inside an impenetrable cavity of adjustable radius, 
whereby the electrostatic Hamiltonian is modified by adding a confining potential in terms of radius $r_c$. 
Ever since the seminal work on confined hydrogen atom (CHA) placed inside a hard spherical cage \cite{michels37}, 
this prototypical model was vigorously studied in reference to the eigenspectra, degeneracy pattern, static and 
dipole polarizability, nuclear magnetic screening constant, excited-state life time, pressure, hyperfine 
splitting constant, filling of electronic shells and many others. \emph{Exact} analytical solution of CHA was 
reported \cite{burrows06} in terms of Kummer $M$-function (confluent hypergeometric). A huge amount of literature 
exists on the topic; they could be found from the references cited. 

Confinement in \emph{many-electron atom} becomes challenging due to the obvious presence of electron-electron Coulomb 
interaction, which breaks the $O(4)$ and $SU(3)$ symmetries of simplified one-electron system. This has generated 
considerable interest and curiosity to investigate the properties of a He atom centrally placed in a hard, rigid 
spherical cage. It serves as a precursor to further atomic confinement studies. Energies as functions of $r_c$ for 
a compressed He centered in a spherical impenetrable cavity was reported as early as in 1952 by a variational 
calculation \cite{tenseldam52}. Thereafter Roothaan Hartree-Fock (RHF) calculation with Slater-type basis 
functions \cite{ludena78}, configuration interaction (CI) \cite{ludena79,rivelino01}, quantum Monte 
Carlo \cite{joslin92}, direct variational \cite{marin91, banerjee06, flores-riveros10, lesech11}, B-splines method 
\cite{ting-yun01}, perturbation theory \cite{flores-riveros10,montgomery10}, explicitly correlated Hylleraas-type 
wave functions within variational framework \cite{aquino03, flores-riveros08,laughlin09, wilson10, montgomery13, 
bhattacharyya13, montgomery15, saha16}, a combination of quantum genetic algorithm and HF method \cite{yakar11}, 
variational Monte Carlo \cite{doma12,sarsa11}, HF calculation employing local and global basis sets \cite{young16},
were adopted to produce better ground-state energies, ionization energies, critical cage, polarizability, 
hyperpolarizability, etc. Apart from the ground state, some low-lying excited 
states were also investigated. They include energies, polarizabilities and other properties in 1sns 
$^{1,3}$S \cite{banerjee06,flores-riveros08,flores-riveros10, yakar11, sarsa11,montgomery13, 
bhattacharyya13, montgomery15, saha16, pupyshev17}, 1s2p $^{1,3}$P \cite{banerjee06, yakar11, pupyshev17}, 
1s3d $^{1,3}$D, and some doubly excited states \cite{yakar11,pupyshev17}. Note that spectroscopic properties of atoms under such 
spatially restricted environment not only involves ground state, but also essentially requires to probe the role 
played by excited states. 

Similar to the confinement in one- and two-electron atoms, such studies were undertaken in other atoms in 
periodic table as well. Thus energy spectra of ground and excited states, as well as filling of shells in 
multi-electron atoms (more than two electrons) was investigated by means of a RHF-type calculation \cite{ludena78,garza12},
configuration average HF \cite{connerade00}, HF \cite{patil05}, variational Monte Carlo 
\cite{sarsa11}, parameterized optimized effective potential (POEP) method using an appropriate cut-off factor 
\cite{sarsa14}, B-spline random phase approximation with exchange \cite{ludlow15}, explicitly correlated and 
multi-configuration variational method \cite{galvez17} and so on. The effect of confinement on correlation energy 
in case of atoms of first few rows were discussed \cite{sabin09, sen14, sanchez16, vyboishchikov16, galvez17}. 
Behavioral changes of compressed atoms under such very tight positions have been comprehensively reviewed by 
several researchers \cite{buchachenko01, dolmatov04, grochala07}. 
 
Apart from the methods mentioned above, there were attempts to pursue the problem through an alternate density
functional theory (DFT). The ground states of a many-electron atom trapped in a 
spherical cavity was treated within an exchange-only framework (two functionals, \emph{viz.}, local 
density approximation (LDA) \cite{parr89} and Becke-88 exchange potential \cite{becke88a} were adopted); the radial 
Kohn-Sham (KS) equation was solved satisfying the Dirichlet boundary condition via numerical shooting method 
\cite{garza98}. Later, ground and 1s2s ($^3$S, $^1$S) states of confined He were presented \cite{aquino06} by 
taking into consideration of LDA exchange-correlation (XC) (Perdew-Wang parametrization for correlation 
\cite{perdew92}), with and without self-interaction correction (SIC). Reactivity indices
like, electronegativity, global hardness, softness, HOMO-LUMO gap were been considered for ground state of 
such systems \cite{garza05} using same XC potential. Similar study was performed by engaging Perdew-Burke-Ernzerhof
(PBE) functional \cite{burgoo08}. A DFT-based variation-perturbation approach 
\cite{waugh10} was proposed to calculate polarizability and hyperpolarizability in confined He. Lately, they were studied 
\cite{vyboishchikov15} via local exchange potentials approximated by Zhao-Morrison-Parr (ZMP) and 
Becke-Johnson (BJ) model. A correlation functional was designed involving the \emph{ab initio} correlation 
energy density for confined atoms \cite{vyboishchikov17}. One-parameter hybrid exchange functional in the form of
PBE exchange coupled with exact-exchange is applied for closed shell atoms inside penetrable and impenetrable 
cavity \cite{alcaraz19}. Recently, a DFT-based study was reported for confined first row transition metal cations 
\cite{espinosa16}. Confinements within \emph{penetrable} walls were undertaken \cite{sanchez16} within a basis-set DFT 
for LDA and generalized-gradient approximated XC functionals. Apart from that, bonding, reactivity and dynamics of an atom 
encapsulated in fullerene cage is also investigated \cite{chakraborty19}. This discussion clearly suggests that, excluding 
a few cases, DFT calculations for confined atoms are restricted typically to ground state.    
  
In this work, our objective is to present a general time-independent DFT scheme which can be applied for both 
ground as well as excited state of an atom, at any given radius of cavity. This is accomplished by 
invoking a simple, physically motivated work-function-based exchange potential \cite{sahni90, sahni92}. 
In order to include correlation effects, 
we adapt two correlation functionals, \emph{viz.,} a simple, local, parametrized Wigner-type \cite{brual78} and the other 
one, a slightly involved nonlinear Lee-Yang-Parr (LYP) \cite{lee88} functional. This procedure was very successfully applied to a 
large number of excited states in \emph{free} atoms, such as single, double, triple excitation; low- and high-lying excitation; 
valence as well as core excitation; hollow and satellite states; autoionizing and high-lying Rydberg states \cite{singh96,roy97,
roy97a,roy02,roy04,roy05}. With this choice of XC functionals, the resultant KS equation is solved using the generalized pseudospectral
(GPS) method imposing the appropriate boundary condition. Recently this has been successfully applied to estimate Shannon entropy 
in some low-lying states of confined He-isoelectronic series \cite{majumdar20}. In this work, we focus on the so-called \emph{hard} 
confinement of an arbitrary atom 
within a spherical rigid cavity, defined by radius $r_c$. This prescription is applicable to \emph{soft or penetrable} boundary as well. 
A detailed systematic results on energy, density and selected expectation values are provided for ground and low-lying 
singly excited 1s2s $^{3,1}$S, 1s2p $^{3,1}$P, 1s3d $^{3,1}$D states of He; lowest states of Li$^+$, Be$^{2+}$, as well as
Li and Be atom. In order to 
estimate the effects of correlation as well as to assess the performance of correlation energy functionals, both X-only and correlated 
results are given. Converged results are offered for low, intermediate and large $r_c$. Section~II 
outlines the methodology used, Sec.~III discusses the results along with comparison with literature, while Sec.~IV draws 
a few conclusions.  

\section {Methodology}
In this section, we first briefly outline the work-function methodology for an arbitrary state corresponding to
an electronic configuration of a given atom. Then we discuss the GPS scheme employed for solution of target KS 
equation. Our starting point is the single-particle time-independent KS equation with imposed confinement, 
conveniently written as, 
\begin{equation}
\label{eq:1}
    \Hvec(\rvec)\psi_i(\rvec)=\epsilon_{i}(\rvec)\psi_{i}(\rvec) ,
\end{equation}
where $\Hvec$ denotes the effective KS Hamiltonian, given by,
\begin{eqnarray}
\label{eq:2}
    \Hvec(\rvec) & = & -\frac{1}{2}\nabla^{2}+v_{eff}(\rvec) \nonumber \\
    v_{eff}(\rvec)& = & v_{ne}(\rvec) +\int \frac{\rho(\rvec^{\prime})}{|\rvec - \rvec^{\prime}|}
 \mathrm{d}\rvec^{\prime}+\frac{\delta E_{xc}[\rho(\rvec)]}{\delta \rho(\rvec)} + v_{conf}(\rvec) .  
\end{eqnarray}
Here $v_{ne}(\rvec)$ signifies the external potential, whereas second and third terms in right-hand side represent 
classical Coulomb (Hartree) repulsion and XC potentials respectively. We assume here that the atom is placed at the 
center of a cavity of radius $r_c$ with impenetrable surface which may be described by a potential of the form,
\begin{equation} v_{conf} (\rvec) = \begin{cases}
0,  \ \ \ \ \ \ \ r \leq r_{c}   \\
+\infty, \ \ \ \  r > r_{c}.  \\
 \end{cases} 
\end{equation}

Though DFT has achieved impressive success in explaining the electronic structure and properties of many-electron system 
in their ground state, calculation of excited-state energies and densities has remained somehow problematic. This is 
mainly due to the absence of a Hohenberg-Kohn theorem parallel to ground state, as well as the lack of a suitable XC 
functional valid for a general excited state. In this work, we have employed an accurate work-function-based exchange potential, 
$v_x(\rvec)$, which is physically motivated. Accordingly, exchange energy can be interpreted as an interaction energy between an 
electron at $\rvec$ and its Fermi-Coulomb hole 
charge density $\rho_{x}(\rvec,\rvec^{\prime})$ at $\rvec^{\prime}$ and given by \cite{sahni90, sahni92}, 
\begin{equation}
\label{eq:4}
 E_{x}[\rho(\rvec)] = \frac{1}{2}\int\int\frac{\rho(\rvec)\rho_{x}(\rvec,\rvec^{\prime})}{|\rvec - \rvec^{\prime}|}
\mathrm{d}\rvec\mathrm{d}\rvec^{\prime}
\end{equation}
The assumption is such that a unique local potential exists for a particular state characterized by usual 
quantum numbers $n,l,m$. One can then define $v_{x} (\rvec)$ as the work done in bringing an 
electron to the point $\rvec$ against the electric field, $\mathcal{E}_{x}(\rvec)$, produced by its Fermi-Coulomb hole 
density, and can be expressed as,  
\begin{equation}
\label{eq:5}
v_{x} (\rvec)  = -\int_{\infty}^{r} \mathcal{E}_{x}(\rvec) \textbf{.} \mathrm{d} \lvec , 
\end{equation}
where
\begin{equation}
\label{eq:6}
\mathcal{E}_{x}(\rvec) = \int\frac{\rho_{x}(\rvec,\rvec^{\prime})(\rvec - \rvec^{\prime})}
{|\rvec - \rvec^{\prime}|^{3}} \mathrm{d}{\rvec^{'}}.
\end{equation}
The Fermi hole charge distribution is known precisely in terms of orbitals,
\begin{equation}
\label{eq:7}
\rho_{x}(\rvec,\rvec^{\prime})=-\frac{\left|\gamma(\rvec,\rvec^{\prime})\right|^2}{2 \rho(\rvec)},
\end{equation}
and consequently the potential can be determined accurately. Here $\left|\gamma(\rvec,\rvec^{\prime})
\right|=\sum_{i} \phi_{i}^{*}(\rvec) \phi_{i}(\rvec^{\prime})$, is single-particle density matrix, 
$\phi_{i}(\rvec)$ is single-particle orbital and $\rho(\rvec)$ is electron density. Note that the potential 
does not have a definite functional form; the respective electronic configuration corresponding to a particular state 
defines it. In this sense this is universal, as the same equation is now valid for both ground and excited states. 
Now, the electron density can be obtained in terms of occupied orbitals as,
\begin{equation}
\label{eq:8}
 \rho(\rvec) = \sum_{i=1}^{N} |\psi_i(\rvec)|^2.  
\end{equation}
For spherically symmetric systems, Eq.~\ref{eq:6} can be simplified in the form \cite{sahni90,sahni92},
\begin{equation}
\label{eq:9}
\begin{aligned}
\mathcal{E}_{x,r}(r) & = {}   \frac{1}{2\pi\rho(r)} \int \sum_{n,l,m,n',l',m',l''} R_{nl}(r) R_{nl}(r')R_{n'l'}(r) 
    R_{n'l'}(r')  
   \left[\frac{\partial }{\partial r} \frac{r^{l''}_{<}}{{r}^{l''+1}_{>}}\right] \\
 & \times r'^2 \mathrm{d}r' \frac{2l+1}{2l'+1} 
  \times C^{2} \left(l l'' l';m, m'-m,m'\right) C^{2} \left(l l'' l';000\right),  
\end{aligned}
\end{equation}
where $R_{nl}(r)$ denotes the radial part of orbitals, and C's signify Clebsch-Gordon coefficients.

It should be noted that this is an orbital dependent, non-variational method, which is applicable for both ground 
and excited states. The nagging orthogonality requirement of a given excited state with all other lower states of same 
space-spin symmetry is indirectly bypassed. In this way, while $v_x (\rvec)$ is incorporated correctly, one 
needs to approximate the correlation potentials. In present calculation, we have used simple Wigner \cite{brual78} 
and slightly involved LYP \cite{lee88} energy functionals to include correlation effects. 

With the above choice of $v_{x}(\rvec)$ and $v_{c}(\rvec)$, the KS differential equation needs to be solved 
numerically maintaining self consistency. For an accurate and efficient solution, we have adopted the GPS method 
leading to a 
non-uniform, optimal spatial discretization. It is computationally orders of magnitude faster than the traditional 
finite-difference methods. The characteristic feature of this method lies in approximating an \emph{exact} function 
$f(x)$, defined in a range $[-1,1]$, by an $N$th-order polynomial $f_{N}(x)$,
\begin{equation} \label{eqn:10}
f(x) \cong f_{N}(x)=\sum_{j=0}^{N} f(x_{j})g_{j}(x),
\end{equation}     
and ensure the estimation to be \emph{exact} at \emph{collocation points} $x_{j}$,
\begin{equation}
\label{eq:11}
f_{N}(x_{j})=f(x_{j}). 
\end{equation}
The radial domain of the atom constrained within a sphere of radius $r_c$ $(r \in [0, r_c])$ is mapped
onto the finite interval $[-1,1]$ via the following algebric nonlinear mapping function, 
\begin{equation}
r=r(x)=L\ \ \frac{1+x}{1-x+\alpha},
\end{equation}
where L and $\alpha=2L/r_{c}$ are two mapping parameters.

Here we employ the Legendre pseudo-spectral method requiring $x_{0}=-1,x_{N}=1$, whereas $x_{j}(j=1,....,N-1)$ are 
defined by the roots of first derivative of Legendre polynomial $P_{N}(x)$, with respect to $x$, namely,
\begin{equation}
\label{eq:12}
P_{N}'(x_{j})=0.
\end{equation}
In Eq.~(\ref{eqn:10}), $g_{j}(x)$ are termed \emph{cardinal functions}, and as such, are given by,
\begin{equation}
\label{eq:13}
g_{j}(x)=-\frac{1}{N(N+1)P_{N}(x_{j})}\frac{(1-x^{2})P_{N}'(x)}{(x-x_{j})},
\end{equation} 
fulfilling the unique property that, $g_{j}(x_{j'})=\delta_{j',j}$. Then use of a non-linear mapping followed by a 
symmetrization procedure, eventually leads to a symmetric eigenvalue problem, which can be solved by readily standard 
softwares to provide accurate eigenvalues and eigenfunctions. The intermediate steps and other details have 
been discussed at length in several publications (see e.g., \cite{roy02,roy04, roy04a,roy04b,roy05} and references
therein); hence they are omitted here.  

Now, our desired confinement is introduced by requiring that the total electron density vanishes at the boundaries 
of spherical cavity that surrounds an atom, which is achieved by imposing the boundary condition, 
$\psi_{nl}(0) = \psi_{nl}(r_c) = 0.$ The self-consistent orbitals can be used to construct relevant Slater determinants 
corresponding to a specific electronic configuration, from which multiplet energies can be estimated following 
Slater's diagonal sum rule \cite{ziegler77,singh96,roy97,roy97a,roy02,roy04,roy05}. 

It is worth mentioning here that, in literature, majority works deal with moderate to large $r_c$ region;  
much fewer papers have been devoted to smaller $r_c$. As, in this work, we have not faced any extra difficulty
to obtain converged solution in latter scenario, we are able to consider all strengths of confinement with uniform 
accuracy and ease, quite comfortably. A general convergence criteria in energy (10$^{-6}$) and potential (10$^{-5}$) during 
the iterative process, as well as GPS parameters (radial grid point, $n_r =300, L=1$) were employed throughout 
the whole confinement region, and for all states undertaken here. 

\section{Result and Discussion}  
In the following, non-relativistic energies, radial expectation 
values and radial densities will be reported for ground and low-lying singly excited 1s2s $^{3,1}$S, 
1s2p $^{3,1}$P, 1s3d $^{3,1}$D states of a confined He atom. Then these are extended for the ground states of 
He-isoelectronic series ($Z=2-4$), as well as Li and Be atom. 
All results are given in atomic units, unless stated otherwise. To put things in proper perspective, three sets of 
energies are offered, \emph{viz.}, (i) exchange-only (ii) considering Wigner correlation (iii) including LYP correlation. 
Throughout the discussion, these are abbreviated as X-only, XC-Wigner and XC-LYP. Some of these states (especially first few 
low-lying ones) are amongst the heavily studied cases for confined many-electron atoms. Consequently many authors 
have studied them and these are quoted wherever feasible.

\begingroup      
\squeezetable
\begin{table}
\caption {\label{tab:table1}Ground-state energy of radially confined He for different $r_c$. See text for details.}
\centering 
\begin{ruledtabular} 
\begin{tabular}{ c|cc|ccc }
  $r_c$    &   X-only     &  Literature & XC-Wigner & XC-LYP  & Literature \\
\hline
 0.1     & 906.61645   &                          & 906.44963 & 907.45572 &  906.6575\footnotemark[19] \\
 0.2     & 206.20456   &                          & 206.06251 & 206.67838 &   206.1696\footnotemark[19] \\
 0.5     & 22.79096    & 22.79095\footnotemark[1],22.79096\footnotemark[16], 
         & 22.69096    & 22.95926  &  22.7437\footnotemark[3],22.7413\footnotemark[13],22.741303\footnotemark[14],23.099\footnotemark[18],  \\
         &             & 23.32202\footnotemark[17],22.80168\footnotemark[23] 
         &             &           &  22.7423\footnotemark[19],22.70203\footnotemark[21]        \\
 0.6     & 13.36683    & 13.36682\footnotemark[1],13.36683\footnotemark[16],
         & 13.27536    & 13.49338  &  13.3204\footnotemark[3],13.318340\footnotemark[8],13.3343\footnotemark[9],13.3182\footnotemark[13],  \\
         &             & 13.81792\footnotemark[17],13.36759\footnotemark[23]                                                                                 
         &             &           & 13.605\footnotemark[18],13.3186\footnotemark[19],13.31194\footnotemark[21]  \\
 0.8     & 4.65737     & 4.65737\footnotemark[1],4.6610\footnotemark[2],4.65736\footnotemark[16],
         & 4.57870     & 4.72898   &  4.6225\footnotemark[2],4.6125\footnotemark[3],4.610554\footnotemark[8],4.6157\footnotemark[9],4.6104\footnotemark[13], \\
         &             & 5.00895\footnotemark[17],4.65760\footnotemark[23]
         &             &           & 4.8133\footnotemark[18],4.6105\footnotemark[19],4.640665\footnotemark[21]          \\
 1       & 1.06121     & 1.06122\footnotemark[1],1.0625\footnotemark[2],1.354\footnotemark[5],
         & 0.99159     & 1.09882   &  1.0186\footnotemark[2],1.0176\footnotemark[3],1.0142\footnotemark[4],1.015870\footnotemark[8],1.0183\footnotemark[9], \\
         &             & 1.112\footnotemark[6],1.06120\footnotemark[15]$^,$\footnotemark[16]$^,$\footnotemark[23],1.35362\footnotemark[17]  
 & & &   1.0158\footnotemark[13]$^,\footnotemark[19]$,1.015755\footnotemark[14]$^,\footnotemark[15]$,1.17040\footnotemark[18],1.01690\footnotemark[20] \\
 1.2     & $-$0.66461  & $-$0.66461\footnotemark[1],$-$0.6639\footnotemark[2],   & $-$0.72758  & $-$0.64954 & $-$0.7075\footnotemark[2],$-$0.7070\footnotemark[3],$-$0.708716\footnotemark[8],$-$0.7079\footnotemark[9],$-$0.7088\footnotemark[13], \\
  &  & $-$0.66458\footnotemark[23] &  &  & $-$0.7087\footnotemark[19],$-$0.70747\footnotemark[20],$-$0.708609\footnotemark[21],$-$0.708801\footnotemark[22] \\
 1.4     & $-$1.57416  & $-$1.57417\footnotemark[1]$^,\footnotemark[23]$,$-$1.5741\footnotemark[2]
   & $-$1.63213 & $-$1.57468 & $-$1.6151\footnotemark[2],$-$1.6156\footnotemark[3],$-$1.6011\footnotemark[7],$-$1.617154\footnotemark[8],$-$1.6167\footnotemark[9], \\
 &  &    &   &   & $-$1.6173\footnotemark[13],$-$1.6172\footnotemark[19],$-$1.61569\footnotemark[20],$-$1.616875\footnotemark[21]  \\
 1.5     & $-$1.86422  & $-$1.8642\footnotemark[2],$-$1.86422\footnotemark[16],
  & $-$1.92015 & $-$1.87072 & $-$1.9040\footnotemark[2],$-$1.9081\footnotemark[4],$-$1.6914\footnotemark[7],$-$1.906740\footnotemark[8],$-$1.9061\footnotemark[9],  \\
  & & $-$1.64897\footnotemark[17] & & & $-$1.81185\footnotemark[18],$-$1.9067\footnotemark[19],$-$1.90599\footnotemark[20],$-$1.906956\footnotemark[22] \\
 2       & $-$2.56256  & $-$2.56253\footnotemark[1],$-$2.5594\footnotemark[2],$-$2.384\footnotemark[5], & $-$2.61154  & 
$-$2.58790 & $-$2.5977\footnotemark[2],$-$2.6026\footnotemark[3],$-$2.6051\footnotemark[4],$-$2.5028\footnotemark[7],$-$2.603630\footnotemark[8], \\
         &             & $-$2.542\footnotemark[6],$-$2.56258\footnotemark[10],
   &  &  & $-$2.5998\footnotemark[9],$-$2.60403\footnotemark[11]$^,\footnotemark[14]$,$-$2.62589\footnotemark[12],$-$2.6040\footnotemark[13],    \\
         &             & $-$2.56257\footnotemark[16]$^,\footnotemark[23]$,$-$2.39363\footnotemark[17]
   & &   & $-$2.53480\footnotemark[18],$-$2.6036\footnotemark[19],$-$2.60223\footnotemark[20],$-$2.604038\footnotemark[22]    \\ 
 3       & $-$2.83103  & $-$2.83083\footnotemark[1],$-$2.8232\footnotemark[2],$-$2.682\footnotemark[5],
  & $-$2.87460  & $-$2.86888 & $-$2.8652\footnotemark[2],$-$2.8708\footnotemark[3],$-$2.8727\footnotemark[4],$-$2.8684\footnotemark[7],$-$2.871808\footnotemark[8], \\
         &             & $-$2.826\footnotemark[6],$-$2.83105\footnotemark[10],$-$2.83078\footnotemark[16],
  &  &  & $-$2.8636\footnotemark[9],$-$2.87426\footnotemark[11],$-$2.89038\footnotemark[12],$-$2.8725\footnotemark[13],  \\
         &             & $-$2.68201\footnotemark[17],$-$2.83099\footnotemark[23]
         &             &  & $-$2.872495\footnotemark[14],$-$2.82256\footnotemark[18],$-$2.8718\footnotemark[19],$-$2.872494\footnotemark[22]   \\ 
 4       & $-$2.85856  & $-$2.85852\footnotemark[1],$-$2.8537\footnotemark[2],$-$2.718\footnotemark[5], 
    & $-$2.90093  & $-$2.89939 & $-$2.8956\footnotemark[2],$-$2.8988\footnotemark[3],$-$2.9003\footnotemark[4],$-$2.899687\footnotemark[8],$-$2.8931\footnotemark[9], \\
   &  & $-$2.859\footnotemark[6],$-$2.85859\footnotemark[10],$-$2.85854\footnotemark[16],
   &  &  & $-$2.90042\footnotemark[11],$-$2.91691\footnotemark[12],$-$2.9004\footnotemark[13],$-$2.900485\footnotemark[14],$-$2.85552\footnotemark[18],  \\
         &             & $-$2.71807\footnotemark[17],$-$2.85834\footnotemark[23]
   &    &    & $-$2.8997\footnotemark[19],$-$2.89834\footnotemark[20],$-$2.894997\footnotemark[21],$-$2.900486\footnotemark[22]       \\ 
 5       & $-$2.86136  & $-$2.86134\footnotemark[1],$-$2.8589\footnotemark[2],$-$2.723\footnotemark[5],
         & $-$2.90352  & $-$2.90332 & $-$2.9004\footnotemark[2],$-$2.9020\footnotemark[3],$-$2.9032\footnotemark[4],$-$2.8764\footnotemark[7],$-$2.8978\footnotemark[9], \\
         &             & $-$2.863\footnotemark[6],$-$2.86139\footnotemark[10]$^,\footnotemark[15]$,$-$2.86138\footnotemark[16],
 &  &    &  $-$2.90337\footnotemark[11],$-$2.91951\footnotemark[12],$-$2.9034\footnotemark[13],$-$2.903410\footnotemark[14]$^,\footnotemark[22]$, \\
         &             & $-$2.72288\footnotemark[17],$-$2.86129\footnotemark[23]
  &  &   & $-$2.903409\footnotemark[15],$-$2.85856\footnotemark[18],$-$2.9028\footnotemark[19],$-$2.903886\footnotemark[21]         \\ 
 6       & $-$2.86162  & $-$2.86151\footnotemark[1],$-$2.724\footnotemark[5],$-$2.863\footnotemark[6],
         & $-$2.90376  & $-$2.90422 & $-$2.9024\footnotemark[3],$-$2.9035\footnotemark[4],$-$2.903278\footnotemark[8],$-$2.8990\footnotemark[9],   \\
         &             & $-$2.86165\footnotemark[10],$-$2.86162\footnotemark[16],
   & &  & $-$2.90368\footnotemark[11],$-$2.91975\footnotemark[12],$-$2.9037\footnotemark[13],$-$2.903696\footnotemark[14],  \\
         &             & $-$2.72346\footnotemark[17] 
 & & &  $-$2.86000\footnotemark[18],$-$2.9033\footnotemark[19],$-$2.90190\footnotemark[20],$-$2.903460\footnotemark[21]    \\ 
$\infty$ & $-$2.86164  & $-$2.86165\footnotemark[1],$-$2.8615\footnotemark[2], 
         & $-$2.90378  & $-$2.90644 & $-$2.9024\footnotemark[2],$-$2.9025\footnotemark[3],$-$2.9037\footnotemark[4],$-$2.8764\footnotemark[7],   \\
         &             & $-$2.861680\footnotemark[15],$-$2.86168\footnotemark[10]$^,$\footnotemark[23] 
 &  &   &  $-$2.903513\footnotemark[8],$-$2.8999\footnotemark[9],$-$2.90372\footnotemark[11],$-$2.91977\footnotemark[12],     \\
  &  & &  &  & $-$2.9037\footnotemark[13]$^,\footnotemark[19]$,$-$2.903724\footnotemark[14]$^,\footnotemark[15]^,\footnotemark[22]$,$-$2.90201\footnotemark[20]       \\
\end{tabular}
\end{ruledtabular}
\begin{tabbing}
$^{\mathrm{a}}$Ref.~\cite{ludena78}. \hspace{25pt}  \=
$^{\mathrm{b}}$Ref.~\cite{gimarc67}. \hspace{25pt}  \=  
$^{\mathrm{c}}$Ref.~\cite{ludena79}. \hspace{25pt}  \=  
$^{\mathrm{d}}$Ref.~\cite{joslin92}. \hspace{25pt}  \=  
$^{\mathrm{e}}$LSDA result of \cite{garza98}. \hspace{25pt}  \=  
$^{\mathrm{f}}$B88 result of \cite{garza98}.   \\
$^{\mathrm{g}}$Ref.~\cite{rivelino01}. \hspace{25pt}  \=  
$^{\mathrm{h}}$Ref.~\cite{aquino03}. \hspace{25pt}  \=  
$^{\mathrm{i}}$Ref.~\cite{banerjee06}. \hspace{25pt}  \=  
$^{\mathrm{j}}$X-only, LDA-SIC result of \cite{aquino06}. \hspace{25pt}  \=  
$^{\mathrm{k}}$Variational result of \cite{aquino06}.   \\  
$^{\mathrm{l}}$LDA-SIC result of \cite{aquino06}. \hspace{40pt}  \=  
$^{\mathrm{m}}$Ref.~\cite{flores-riveros08}. \hspace{25pt}  \=  
$^{\mathrm{n}}$Ref.~\cite{laughlin09}. \hspace{25pt}  \=  
$^{\mathrm{o}}$Ref.~\cite{wilson10}. \hspace{25pt}  \=  
$^{\mathrm{p}}$Exact exchange result of \cite{waugh10}.  \\  
$^{\mathrm{q}}$X-only LDA result of \cite{waugh10}. \hspace{25pt} \=
$^{\mathrm{r}}$XC-LDA result of \cite{waugh10}. \hspace{25pt} \=
$^{\mathrm{s}}$Ref.~\cite{flores-riveros10}. \hspace{25pt}  \=  
$^{\mathrm{t}}$Ref.~\cite{lesech11}. \hspace{25pt}  \=  
$^{\mathrm{u}}$Ref.~\cite{doma12}.   \\  
$^{\mathrm{v}}$Ref.~\cite{bhattacharyya13}. \hspace{25pt}  \=  
$^{\mathrm{w}}$Ref.~\cite{yakar11}. \hspace{25pt}  \=  
\end{tabbing}
\end{table}
\endgroup 


\subsection{Confined He atom}
Let us first discuss ground-state energies of confined He in Table~\ref{tab:table1}, for 15 representative 
$r_c$'s, starting from a very strong confinement in $r_c=0.1$ to a large $r_c$, corresponding to \emph{free} atom. 
This is the most vigorously studied case, as it bears relevance to the simplest prototypical many-electron atom. 
The X-only results are compared with available literature by tabulating them in column 3.  
The RHF energies \cite{ludena78} within a Clementi-Roetti basis, are found for all $r_c$ except 0.1, 0.2, 1.5. 
For the entire region of confinement, present results of column 2 show excellent agreement. Note that, in several 
instances, the two energies are identical. A similar agreement is also noticed for the exact exchange result 
(denoted by EXX in Table~\ref{tab:table1} of \cite{waugh10}) throughout the entire $r_c$. Apart from these two cases, 
another reference pertaining to this case includes HF \cite{gimarc67,wilson10}. The X-only result 
(XO-LDA of \cite{waugh10}) employing local Dirac exchange functional \cite{dirac30}, are also linked in
present scenario. Generally, these energies remain consistently above our X-only result. The discrepancy is more 
prominent in lower $r_c$'s, with absolute deviation hovering within the range 2.3-21.6\%. Similar calculations within LDA 
by various researchers \cite{garza98, aquino06, waugh10} are also referred. The inclusion of SIC in LDA X-only framework 
\cite{aquino06} improves energies and at larger $r_c$, it yields values close to HF. Present X-only scheme performs 
better compared to LDA. A GGA-based DFT \cite{garza98} using B88 functional \cite{becke88a}, provides
better approximation to HF than LDA. Our scheme provides appreciable agreement with GGA results. It is necessary 
to mention that, at strong confinement ($r_c < 4$) region these GGA energies are higher relative to the present case. 
However, at large $r_{c}$, a reverse situation is noticed. A slightly compromised matching is observed with 
\cite{yakar11}, where a combination of quantum genetic algorithm and RHF is utilized.   

\begingroup           
\squeezetable
\begin{table}
\caption {\label{tab:table2}Energy values of 1s2s $^{3,1}$S states of He confined in a spherical cavity of radius $r_c$.}
\begin{ruledtabular} 
\begin{tabular}{c|c|c c c|c|c c c}
& \multicolumn{4}{c|}{$^3$S} & \multicolumn{4}{c}{$^1$S} \\
\cline{2-9}
$r_c$  & X-only     & XC-Wigner   & XC-LYP     & Literature  &  X-only    & XC-Wigner    & XC-LYP   & Literature \\
\hline
0.1    & 2370.7389  & 2370.5729   & 2372.1569  & 2388.7273\footnotemark[2],2370.8673\footnotemark[3]
       & 2376.4826  & 2376.3166   & 2377.9005  & 1995.2692\footnotemark[2],2376.8368\footnotemark[3]              \\
0.2    & 568.19924  & 568.05832   & 569.03586  & 572.3488\footnotemark[2],568.2066\footnotemark[3] 
       & 571.14431  & 571.00342   & 571.98094  & 485.3350\footnotemark[2],571.1854\footnotemark[3]                \\
0.3    & 241.50286  & 241.37997   & 242.08839  & 243.1949\footnotemark[2],241.5068\footnotemark[3] 
       & 243.51696  & 243.39410   & 244.10248  & 209.3899\footnotemark[2],243.5355\footnotemark[3]                 \\
0.5    & 78.83157   & 78.73300    & 79.17557   & 79.3341\footnotemark[2],78.8352\footnotemark[3]
       & 80.10411   & 80.00552    & 80.44798   & 70.6581\footnotemark[2],80.1154\footnotemark[3]     \\
       & 79.43685\footnotemark[1] &            & & 
       & 79.65350\footnotemark[1] & & &      \\
0.6    & 51.86713   & 51.77718    & 52.14254   & 52.1803\footnotemark[2],51.8709\footnotemark[3]
       & 52.95550   & 52.86571    &  53.23094  & 47.4173\footnotemark[2],52.9658\footnotemark[3]     \\
       & 52.23345\footnotemark[1] &            & & 
       & 52.68074\footnotemark[1] & & &      \\
0.8    & 25.86532   & 25.78841    & 26.04941   & 26.0029\footnotemark[2],25.8693\footnotemark[3]
       & 26.72502   & 26.64834    & 26.90914   & 24.8168\footnotemark[2],26.7338\footnotemark[3]     \\
       & 26.06405\footnotemark[1] &            & & 
       & 26.14839\footnotemark[1] & & &      \\
1      & 14.36896   & 14.30144    & 14.49546   & 15.5451\footnotemark[4],14.3598\footnotemark[5],
       & 15.09248   & 15.02498    & 15.21902   & 14.5358\footnotemark[2],15.0998\footnotemark[3]    \\
       & 14.43401\footnotemark[1] &            &            & 14.3599\footnotemark[2],14.3733\footnotemark[3]
       & 14.87847\footnotemark[1] &            &            &                       \\
1.5    & 3.81468    & 3.76201     & 3.86349    & 3.8068\footnotemark[2],3.8199\footnotemark[3]
       & 4.3546     & 4.30191     & 4.40345    & 4.3045\footnotemark[2],4.3567\footnotemark[3]                       \\
2      & 0.56698    & 0.52281     & 0.57954    & 0.5862\footnotemark[4],0.56026\footnotemark[8], 
       & 1.0021     & 0.95795     & 1.0146     & 0.9639\footnotemark[2],0.9977\footnotemark[3]          \\
       & 0.56961\footnotemark[1],  &            &            
       & 0.58463\footnotemark[9],0.5603\footnotemark[5]$^,\footnotemark[2]$, & 0.60713\footnotemark[1],  &            
       &            &                    \\
       & 0.56698\footnotemark[6]           &             &            & 0.5733\footnotemark[3]
       & 0.99333\footnotemark[6]           &             &            &                    \\
2.5    & $-$0.74592 & $-$0.78479  & $-$0.75220 & $-$0.7516\footnotemark[5],$-$0.751657\footnotemark[7] 
       & $-$0.39500 & $-$0.43469  & $-$0.40158 & $-$0.433214\footnotemark[10]     \\
3      & $-$1.36562 &  $-$1.40091 & $-$1.38218 & $-$1.3679\footnotemark[4],$-$1.37046\footnotemark[8], 
       & $-$1.09026 & $-$1.12545  & $-$1.10734 & $-$1.0991\footnotemark[2],$-$1.0937\footnotemark[3], \\
       & $-$1.36799\footnotemark[1], &    &   & $-$1.34861\footnotemark[9],$-$1.3705\footnotemark[5]$^,\footnotemark[2]$,
       & $-$1.34955\footnotemark[1], &   &   & $-$1.114121\footnotemark[10]     \\
       & $-$1.36605\footnotemark[6]  &  &    & $-$1.3562\footnotemark[3]    & $-$1.08419\footnotemark[6]  &   &   &     \\
4      & $-$1.87095 &  $-$1.90207 & $-$1.89643 & $-$1.8734\footnotemark[4],$-$1.87461\footnotemark[8],
       & $-$1.70881 & $-$1.73987  & $-$1.73465 & $-$1.6949\footnotemark[2],$-$1.6913\footnotemark[3]      \\
       & $-$1.87331\footnotemark[1], &          &            & 
         $-$1.86296\footnotemark[9],$-$1.8746\footnotemark[5]$^,\footnotemark[2]$, & $-$1.86047\footnotemark[1], &          
       &            &         \\
       & $-$1.87162\footnotemark[6]  &   &   & $-$1.8569\footnotemark[3],$-$1.874612\footnotemark[7]  &  $-$1.70025\footnotemark[6] & & & \\
5      & $-$2.04515 & $-$2.07406  & $-$2.07324 & $-$2.0473\footnotemark[4],$-$2.04804\footnotemark[8],
       & $-$1.94379 & $-$1.97274  & $-$1.97136 & $-$1.8684\footnotemark[2],$-$1.9051\footnotemark[3]     \\
       & $-$2.04787\footnotemark[1],  &    &  & $-$2.04317\footnotemark[9],$-$2.0480\footnotemark[5]$^,\footnotemark[2]$,
       & $-$1.94676\footnotemark[1],  &   &   & $-$1.949761\footnotemark[10]          \\
       & $-$2.04590\footnotemark[6]   &             &            & $-$2.0250\footnotemark[3],$-$2.048044\footnotemark[7]
       & $-$1.93612\footnotemark[6]   &             &            &                    \\
6      & $-$2.11539 &  $-$2.14302 & $-$2.14389 & $-$2.1171\footnotemark[4],$-$2.11782\footnotemark[8],
       & $-$2.04408 & $-$2.07167  & $-$2.07084 & $-$1.9136\footnotemark[2],$-$1.9880\footnotemark[3]   \\
       & $-$2.11615\footnotemark[6]       &      &     & $-$2.11609\footnotemark[9],$-$2.1178\footnotemark[5]$^,\footnotemark[2]$,
       & $-$2.04168\footnotemark[6]            &             &            &                   \\
       &            &             &            & $-$2.0880\footnotemark[3] 
       &            &             &            &                    \\
10     & $-$2.17079 &  $-$2.19663 & $-$2.19745 & $-$2.1714\footnotemark[4],$-$2.17262\footnotemark[8],
       & $-$2.12726 & $-$2.15583  & $-$2.15087 & $-$2.139619\footnotemark[10] \\
       & $-$2.17146\footnotemark[1] &          &            & $-$2.17345\footnotemark[9],$-$2.1726\footnotemark[5],
       & $-$2.13092\footnotemark[1] &          &            &                    \\
       &                            &          &            & $-$2.172627\footnotemark[7]
       &                            &          &            &             \\
$\infty$&$-$2.17342 &  $-$2.19990 & $-$2.19918 & $-$2.17523\footnotemark[8],$-$2.1752\footnotemark[5]$^,\footnotemark[2]$, & $-$2.13488 & $-$2.16928  & $-$2.15214 & $-$1.9264\footnotemark[2],$-$2.0364\footnotemark[3],  \\
       & $-$2.17424\footnotemark[6]  &    &    & $-$2.17622\footnotemark[9],$-$2.1241\footnotemark[3] 
       & $-$2.14340\footnotemark[6]            &             &            & $-$2.145974\footnotemark[10]   \\
\end{tabular}
\end{ruledtabular}
\begin{tabbing}
$^{\mathrm{a}}$Ref.~\cite{yakar11}. \hspace{25pt}  \=  
$^{\mathrm{b}}$GH result of \cite{flores-riveros10}. \hspace{25pt}  \= 
$^{\mathrm{c}}$PT result of \cite{flores-riveros10}. \hspace{25pt}  \=  
$^{\mathrm{d}}$Ref.~\cite{banerjee06}. \hspace{15pt}  \=
$^{\mathrm{e}}$Ref.~\cite{flores-riveros08}. \hspace{25pt}  \=  
$^{\mathrm{f}}$Ref.~\cite{luzon20}. \hspace{25pt}  \=   \\
$^{\mathrm{g}}$Ref.~\cite{saha16}. \hspace{25pt}  \=  
$^{\mathrm{h}}$Variational result of \cite{aquino06}. \hspace{25pt}  \=  
$^{\mathrm{i}}$LDA-SIC result of \cite{aquino06}. \hspace{25pt}  \=  
$^{\mathrm{j}}$Ref.~\cite{bhattacharyya13}. \hspace{25pt}  \=   
\end{tabbing}
\end{table}
\endgroup

Columns 4, 5 of Table~\ref{tab:table1} now report XC-Wigner and XC-LYP energies, along with literature results in 
column 6. The differences in two energies remain in the range of 0.0002-1.006 a.u. In free-limit scenario, 
the two energies come quite closer. References are more prevalent for $r_c \geq 0.5$ than $r_c < 0.5$. In general, 
from \emph{free} atom, at $r_c \to \infty$, compression of the box is accompanied by an increase in energy. 
Overall energy comprises two terms, namely, confinement kinetic and Coulomb interaction energy. As an atom is enclosed, 
it gets constrained in a shorter box, leading to a net accumulation of kinetic energy. For all $r_c$'s, energies can be 
compared with that of \cite{flores-riveros10}, involving a wave function expanded via a generalized Hylleraas basis (GHB) 
(10 terms) along with a suitable cut-off factor. Both Wigner and LYP functionals produce reasonably good agreement with these, 
recording absolute deviations of 0.02-9.23\% and 0.01-8.35\%. In strong to moderate zone, energies distinctly remain nearer 
to XC-Wigner; however with an enhancement of box size, three energies eventually tend to approach each other closely. A 25-term 
variational function in a GHB with a cut-off function \cite{flores-riveros08, aquino06} generally gives energies in between 
the two functionals; discrepancies range in between 0.002-2.65\% and 0.003-8.36\%, indicating slight advantage of Wigner 
correlation. Except for first two box sizes and $r_c=1.5$, these were systematically estimated as early as in 1979 in 
\cite{ludena79} using a 41-term CI expansion generated by a 6s4p4d basis. In moderate to smaller box size ($r_c < 3$) these
correlated reference results seem to be reproduced better by Wigner than LYP; after that the three energies tend 
to corroborate each other. The calculated energies are also compared with CI method involving HB \cite{rivelino01,laughlin09}.
 The variational Monte Carlo energies \cite{doma12} (with a set of 10$^6$ points), by and large, 
maintain similar agreement with two correlation functionals, as the previous two, except for $r_c$ of 0.5 
a.u., where it gives an underestimation by about 0.04 a.u. Quantum Monte Carlo energies \cite{joslin92}, 
(for $1 \leq r_c \leq 2.5$), show better consensus with Wigner than LYP. For $r_c \geq 1$, the Rayleigh-Ritz 
calculation \cite{lesech11} also provides similar conclusions as above references, leading to deviations of 0.06-2.84\%, 
0.04-8.19\% for Wigner and LYP. Except for first two $r_c$'s, energies were published \cite{banerjee06} from a 
two-parameter, correlated two-particle variational function; two functionals differ by 0.16-2.78\% and 0.18-8.22\%. 
The KS DFT \cite{waugh10}, with an STO basis, consistently produces higher energies than both XC-Wigner 
(by 1.53-15.28\%) and XC-LYP (by 0.61-6.12\%). The LDA XC energies with SIC published 
in \cite{aquino06} show decent agreement for intermediate to large $r_c$.

\begingroup            
\squeezetable
\begin{table}
\caption{\label{tab:table3} Singly-excited 1s2p $^{3,1}$P and 1s3d $^{3,1}$D energies of He at various $r_c$, in a.u.}
\begin{ruledtabular} 
\begin{tabular}{c|c c|c c c|c c|c c c}
	&       \multicolumn{5}{c}{1s2p $^3$P}     &  \multicolumn{5}{c}{1s2p $^1$P}        \\
	\cline{2-11}
$r_c$ & X-only & Lit.  & XC-Wigner   & XC-LYP  & Lit.   &  X-only & Lit.  & XC-Wigner   & XC-LYP  & Lit. \\
\hline
0.1  & 1429.4156  &             & 1429.2504   &  1430.4981  &                             
     & 1436.5840  &             &  1436.4189  &  1437.6666  &        \\
0.5  & 45.01798   &  45.06318\footnotemark[1]   & 44.9220     &  45.2679    &                            
     & 46.4261    &  46.48432\footnotemark[1]   &  46.3302    &  46.6761    &             \\
0.8  & 13.83774   & 13.84936\footnotemark[1]    & 13.7638     &  13.9646    &  
     &  14.7012   &  14.71138\footnotemark[1]   &  14.6274    &  14.8281    &             \\
1    & 7.18913    &  7.19649\footnotemark[1],   & 7.1246      &  7.2717     &  7.5265\footnotemark[4],   
     &  7.8689    &  7.87081\footnotemark[1],   & 7.8045      &  7.9515     &  8.0312\footnotemark[4],    \\
     &            & 7.18908\footnotemark[2]     &             &             &  7.680\footnotemark[5]  
     &            & 7.86331\footnotemark[2]     &             &             &  7.751\footnotemark[5]      \\
1.6  & 0.69311    & 0.69364\footnotemark[1]     & 0.6453      &  0.7089     &                         
     &  1.0894    &  1.08013\footnotemark[1]    &  1.0417     &  1.1053     &            \\  
1.8  & $-$0.06149 &  $-$0.06022\footnotemark[1] & $-$0.1059   & $-$0.0572   &                            
     &  0.2798    &  0.26860\footnotemark[1]    &  0.2354     &  0.2841     &            \\
2    & $-$0.56852 & $-$0.56750\footnotemark[1], & $-$0.6102   & $-$0.5729   & $-$0.4907\footnotemark[4],   
     & $-$0.2722  & $-$0.42358\footnotemark[1], & $-$0.3139   & $-$0.2767   & $-$0.3414\footnotemark[4],    \\
     &   & $-$0.56854\footnotemark[2]$^,\footnotemark[3]$  &  &             & $-$0.3692\footnotemark[5]
     &   & $-$0.28495\footnotemark[2]$^,\footnotemark[3]$  &  &             & $-$0.3334\footnotemark[5]      \\
5    & $-$2.01896 &  $-$2.01930\footnotemark[1],              & $-$2.0475   & $-$2.0506   & $-$2.0126\footnotemark[4], 
     & $-$1.9631  &  $-$1.99797\footnotemark[1],              & $-$1.9909   & $-$1.9938   & $-$1.9928\footnotemark[4],   \\
     &   &   $-$2.01977\footnotemark[2]$^,\footnotemark[3]$   &             &             & $-$2.0012\footnotemark[5]          
     &   & $-$1.98051\footnotemark[2]$^,\footnotemark[3]$     &             &             & $-$1.9857\footnotemark[5]      \\
7    & $-$2.0991  &  $-$2.10041\footnotemark[1],              & $-$2.1255   & $-$2.1290   & $-$2.0964\footnotemark[4], 
     & $-$2.0677  &  $-$2.11382\footnotemark[1],              & $-$2.0930   & $-$2.0953   & $-$2.0861\footnotemark[4],   \\
     &   &  $-$2.10049\footnotemark[2]$^,\footnotemark[3]$    &             &             & $-$2.0934\footnotemark[5]     
     &   &  $-$2.08193\footnotemark[2]$^,\footnotemark[3]$    &             &             & $-$2.0812\footnotemark[5]      \\
10   & $-$2.12479  &  $-$2.12665\footnotemark[1],             & $-$2.1497   & $-$2.1518   & $-$2.1234\footnotemark[4], 
     & $-$2.1023   &  $-$2.11563\footnotemark[1],             & $-$2.1258   & $-$2.1255   & $-$2.1172\footnotemark[4],   \\
     &   &  $-$2.12644\footnotemark[2]$^,\footnotemark[3]$    &             &             & $-$2.1249\footnotemark[5]              
     &   &  $-$2.11566\footnotemark[2]$^,\footnotemark[3]$    &             &             & $-$2.1146\footnotemark[5]       \\
\hline
         &       \multicolumn{5}{c}{1s3d $^3$D}     &  \multicolumn{5}{c}{1s3d $^1$D}        \\
\hline
0.1      &  2086.2494  &               &  2086.0853  &  2087.5813  &               
         &  2089.0937  &               &  2088.9296  &  2090.4256  &                          \\
0.5      &  72.18395   & 72.30233\footnotemark[1]    &  72.0901    &  72.5138    &                
 	 &  72.7197    & 72.75302\footnotemark[1]    &  72.6260    &  73.0497    &            \\
1        &  14.26227   & 14.28253\footnotemark[1]    &  14.2000    &  14.3867    &              
         &  14.5037    & 14.40101\footnotemark[1]    &  14.4415    &  14.6282    &            \\
2        &   1.33071   & 1.33232\footnotemark[1]     &  1.2910     &  1.3444     &             
         &  1.4151     & 1.37266\footnotemark[1]     &  1.3753     &  1.4289     &             \\ 
2.2      &   0.66672   & 0.66777\footnotemark[1]     &  0.6291     &  0.6716     &    
	 & 0.7361      & 0.73500\footnotemark[1]     &  0.6984     &  0.7407     &            \\
2.6      & $-$0.20070  & $-$0.19987\footnotemark[1]  & $-$0.2350   & $-$0.2079   &  
	 & $-$0.15415  & $-$0.15599\footnotemark[1]  & $-$0.18889  & $-$0.1615   &    \\
3        & $-$0.72396  & $-$0.72362\footnotemark[1]  & $-$0.7562   & $-$0.7385   & 
	 & $-$0.6930   & $-$0.70951\footnotemark[1]  & $-$0.7255   & $-$0.7077   &   \\
5        & $-$1.67291  & $-$1.67325\footnotemark[1]  & $-$1.7001   & $-$1.6975   &  
	 & $-$1.6684   & $-$1.61835\footnotemark[1]  & $-$1.6959   & $-$1.6931   &  \\
7        & $-$1.9037   & $-$1.9038\footnotemark[1]   & $-$1.9290   & $-$1.9272   &   
	 & $-$1.9026   & $-$1.9035\footnotemark[1]   & $-$1.9281   & $-$1.9261   &   \\
10       & $-$2.00707  & $-$1.98617\footnotemark[1]  & $-$2.0310   & $-$2.0218   & 
	 & $-$2.0068   & $-$2.00707\footnotemark[1]  & $-$2.0308   & $-$2.0213   &            \\
\end{tabular}
\end{ruledtabular}
\begin{tabbing}
$^{\mathrm{a}}$Ref.~\cite{yakar11}. \hspace{55pt}  \=  
$^{\mathrm{b}}$Ref.~\cite{sarsa18_cpl}. \hspace{55pt}  \=
$^{\mathrm{c}}$Ref.~\cite{luzon20}. \hspace{55pt}  \= 
$^{\mathrm{d}}$Ref.~\cite{banerjee06}. \hspace{55pt}  \= 
$^{\mathrm{e}}$Ref.~\cite{patil04}.    
\end{tabbing}
\end{table}
\endgroup  
  
Next, we apply this method in case of excited states. This gives an opportunity to test and assess its performance in 
such states under hard confinement. Thus Table~\ref{tab:table2} furnishes energies of singly excited 1s2s $^3$S and $^1$S 
states of a caged-in He atom for a wide range of $r_c$. Reference theoretical results in this situation are not as diverse 
as in previous table; nevertheless a substantial amount exists, which are duly given in footnote. The presentation strategy 
remains same as before. X-only energies for triplet, singlet states are tabulated at different $r_c$'s in columns 2, 6 
respectively, along with RHF result (for $r_c \geq 0.5$) within an STO basis and invoking a genetic algorithm for 
minimization \cite{yakar11}. Triplet energies remain lower by 0.45-0.76\% when radius remains below 2, while they lie 
0.03-0.46\% above reference otherwise. For singlet state, however, current energies are uniformly overestimated 
(0.17-65.1\%) throughout the whole strength. In singlet case, $r_c \geq 2$ region records maximum deviation. This large 
discrepancy is rather unexpected and unusual, as none of our X-only results in preceding (for ground state) and any 
of the future tables (as will be evident from discussion later) produces this kind of differences when 
compared to other accurate theoretical methods. 
Moreover, very recently, a multi-configuration parametrized optimized effective potential (POEP) method \cite{luzon20} has been 
proposed for both these states for $r_c \geq 2$, which offers excellent agreement with X-only energies 
recording 0.001-0.03\% and 0.11-0.88\% of absolute deviations; in a few instances, two energies completely 
merge. Thus, on the light of above facts, we are confident that present energies are better than those of 
\cite{yakar11}; more careful and sophisticated calculations in future will hopefully resolve this issue. Now coming to 
correlated scenario, accurate variational energies \cite{aquino06,flores-riveros08,flores-riveros10,bhattacharyya13} by 
choosing a variety of large expansions in HB are available. These references, more or less, remain in conformity with each other. 
For singlet state, these can be compared with the accurate correlated wave function with 161 terms in HB 
\cite{bhattacharyya13}. Here our energies (both Wigner and LYP) are underestimated, except for XC-LYP at $r_c \leq 3$ where it 
is overestimated. 
Another reference is due a perturbative approach \cite{flores-riveros10}, for both states at all $r_c$'s except at 2.5 
and 10. It records absolute error in ranges of 0.02-3.56\%, 0.05-3.53\% for triplet, while 0.02-6.52\%, 
0.04-5.68\% for singlet, with Wigner and LYP functionals. Triplet energies obtained by solving KS equation with LDA 
and LDA-SIC \cite{aquino06} differ considerably from each other and also from other variational results, for all 
$r_c \geq 2$. 
For higher-$S$ state, accurate energies are also available from Ritz method using an explicitly 
correlated HB \cite{saha16}, for all $r_c \geq 2.18$ a.u. 

\begin{figure}        
\centering
\begin{minipage}[t]{0.48\textwidth}\centering
\includegraphics[scale=0.75]{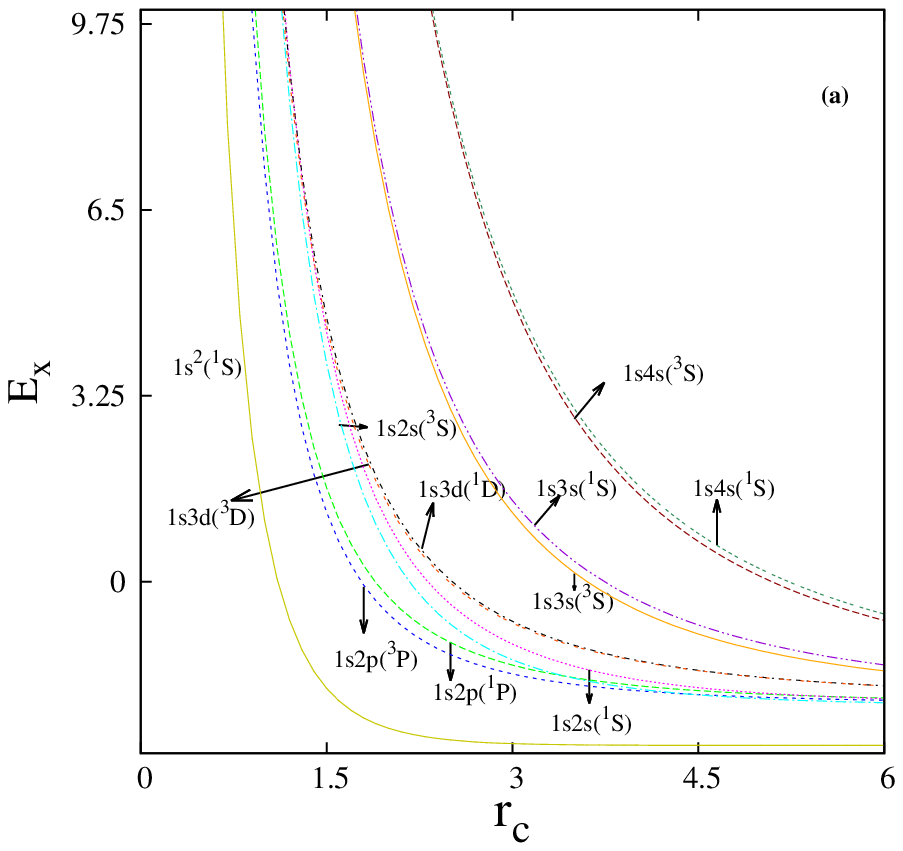}
\end{minipage}
\begin{minipage}[t]{0.48\textwidth}\centering
\includegraphics[scale=0.75]{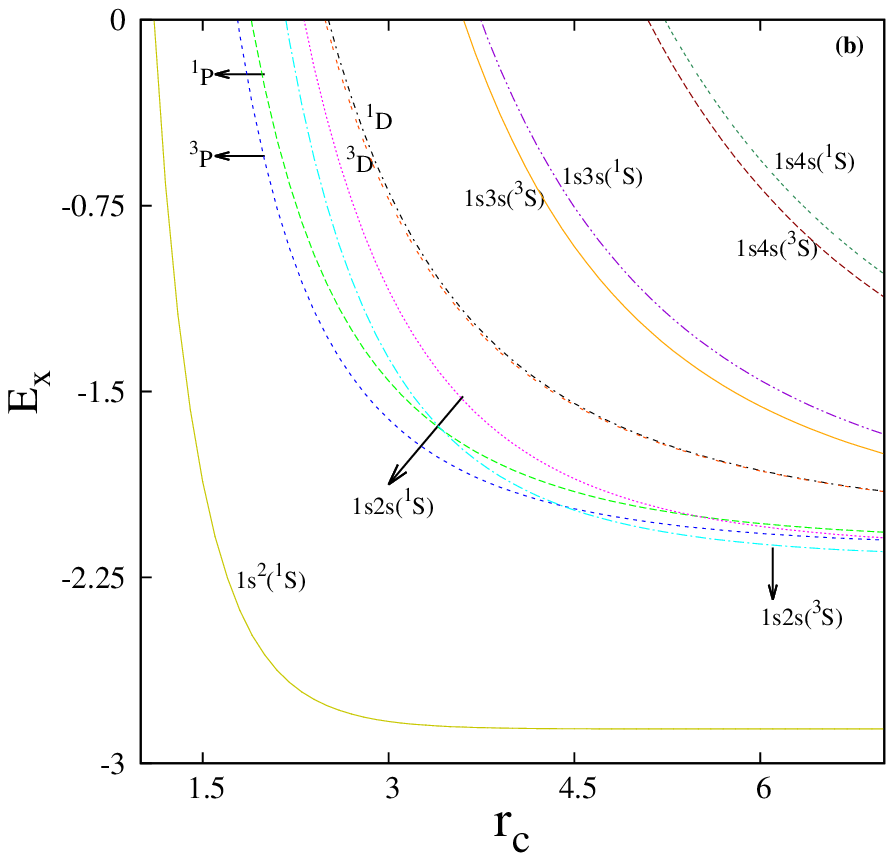}
\end{minipage}
\caption{\label{fig:figure1}Energy changes in some low-lying states of confined He with $r_c$ in (a). Panel (b) shows a 
magnification of (a) in $E \leq 0$ region. See text for details.}
\end{figure} 
  
In order to augment the discussion on excited states, Table~\ref{tab:table3} next offers specimen triplet and singlet 
energies of confined He corresponding to states having total orbital angular momentum $L =1,2$ arising from singly 
excited configurations 1s2p and 1s3d. Maintaining consistency with earlier presentation strategy, X-only results are 
tabulated in 2nd and 7th columns in upper (1s2p) and lower (1s3d) sections, at representative $r_c$'s. As before, 
energies monotonically fall off as the box is enlarged. Literature results in these cases are visibly scarce. For 
$r_c \geq 0.5$, STO-based RHF \cite{yakar11} shows, in general, nice agreement with X-only values for 
entire range, excepting some stray cases where, as in previous table, some spike is observed. Thus, while $^3$P 
remains within 0.02-2.02\%, $^1$P is estimated within 0.02-4.17\%, excepting the special case for 
$r_c= 2$, where an unusually large difference of 35.73\% is recorded. 
For $^{3,1}$P, near-HF energies within a POEP method has been reported lately 
\cite{sarsa18_cpl}, for $r_c \geq 1$. It is encouraging to note that, for $^3$P, our energies match excellently with 
theirs, uniformly for all radii. However, for $^1$P, current energies remain slightly higher when $r_c <5$, 
but steadily lowers as we approach the free-atom unconfined limit. These reference results also fully match with 
\cite{luzon20}, for all available $r_c$. Absolute deviation relative to \cite{sarsa18_cpl} lies in the 
range of 0.001-0.08\% and 0.07-4.47\% for $^3$P, $^1$P. The correlated $^3$P energies are found to be 
lower for both functionals throughout entire strength with respect to the interpolation approach of 
\cite{patil04} and variational scheme \cite{banerjee06}. But for $^1$P, present energies are lower in small $r_c$, 
while producing higher values in moderate to large $r_c$. The energy difference between two 
functionals gradually diminishes as confinement occurs in larger enclosures. For D states, X-only results give 
absolute deviations of 0.005-1.05\% and 0.01-3.1\% for $^3$D and $^1$D. No literature data can be found for 
correlated energies, for direct comparison. Moreover, no DFT calculations are known for any of these states 
in this table.

\begin{figure}                         
\centering
\vspace{1.2cm}
\begin{minipage}[c]{0.48\textwidth}\centering
\includegraphics[scale=0.75]{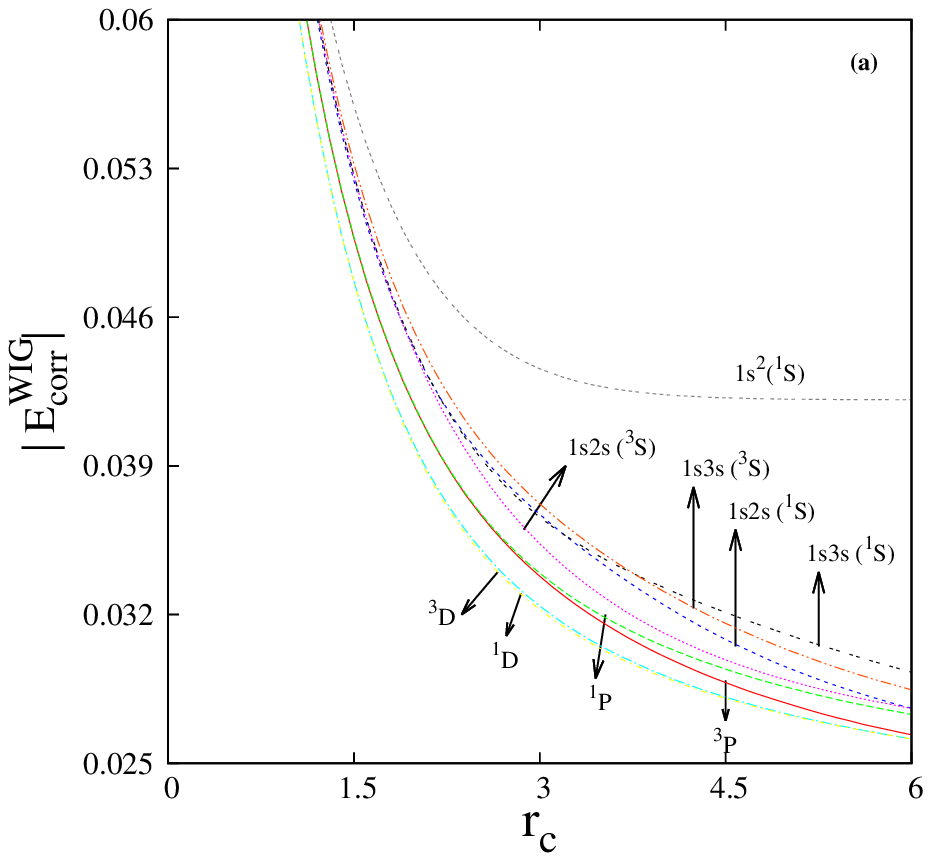}
\end{minipage}%
\begin{minipage}[c]{0.48\textwidth}\centering
\includegraphics[scale=0.75]{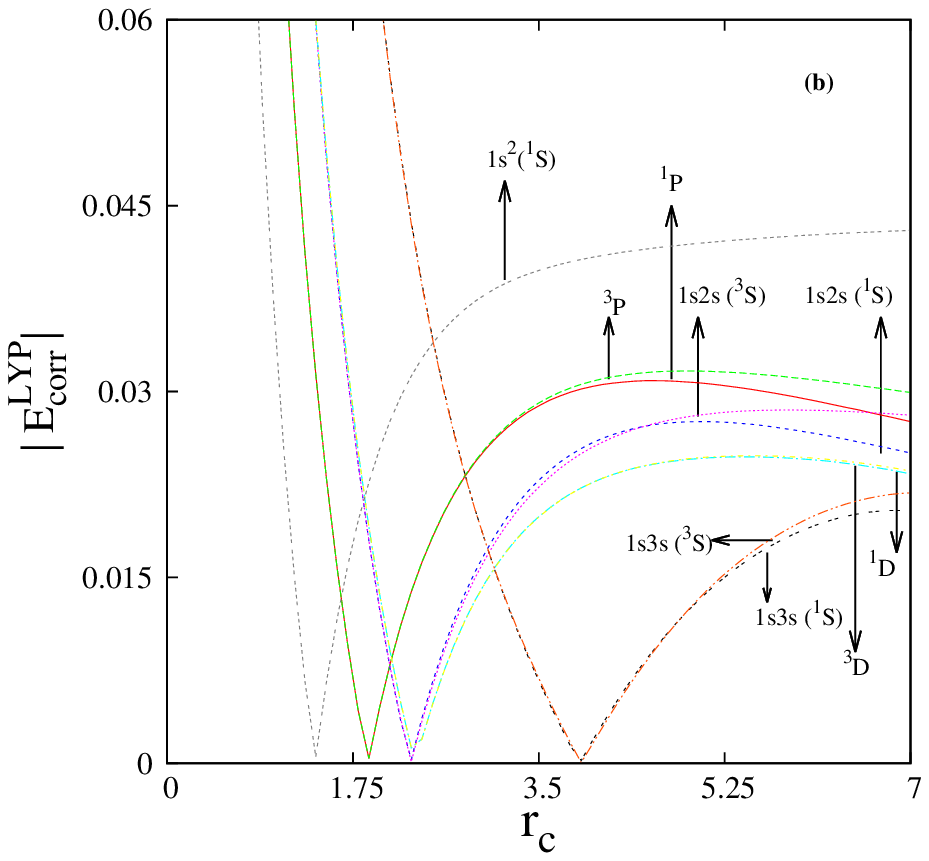}
\end{minipage}%
\caption{\label{fig:figure2}Absolute value of correlation energies for singly excited states of confined He, in panels 
(a) and (b) for Wigner and LYP functionals, with changes in box size. See text for details.}
\end{figure}

In order to illustrate the impact of confinement, energies of compressed He are portrayed in panels (a) and (b) of 
Fig.~\ref{fig:figure1} for a few selected singlet and triplet singly excited states as a function of $r_c$. In addition to 
the states considered in above tables, here we also include 1s3s and 1s4s $^{3,1}$S. 
To get a better understanding of crossing amongst various states, a magnified portion of (a) is displayed in 
(b) in the negative energy region, with improved resolution. Since X-only, XC-Wigner, XC-LYP energies 
produce qualitatively similar plots, we have taken liberty to use X-only energies to illustrate the essential features.
In consonance with discussions of 
Tables~\ref{tab:table1}-\ref{tab:table3}, here also it is evident that, while influence of $v_{conf}(\rvec)$ 
seems to be more effective in smaller $r_c$ in low-lying states, for higher states this impact causes a shift towards 
larger $r_c$. In \emph{free} He atom, the ordering of states under consideration is: E$_{\mathrm{1s4s}(^1\mathrm{S})}>$ 
E$_{\mathrm{1s4s}(^3\mathrm{S})}>$ E$_{\mathrm{1s3d}(^3\mathrm{D})}>$ E$_{\mathrm{1s3d}(^1\mathrm{D})}>$ 
E$_{\mathrm{1s3s}(^1\mathrm{S})}>$ E$_{\mathrm{1s3s}(^3\mathrm{S})}>$ E$_{\mathrm{1s2p}(^1\mathrm{P})}>$ 
E$_{\mathrm{1s2p}(^3\mathrm{P})}>$ E$_{\mathrm{1s2s}(^1\mathrm{S})}>$ E$_{\mathrm{1s2s}(^3\mathrm{S})}>$ 
E$_{\mathrm{1s}^2(^1\mathrm{S})}$. With increase in confinement strength, this ordering gets 
dissolved due to multiple crossing between states, and rearrangement that occurs therefrom; it becomes 
function of $r_c$. Thus at $r_c \approx 5.3$, \textcolor{red}{4.45}, 3.4, crossings occur between 1s2p $^1$P, 1s2s $^1$S; 
1s2p $^{3}$P, 1s2s $^3$S; 1s2p $^{1}$P, 1s2s $^3$S respectively. It is to be mentioned that, beyond the limits 
of $r_c$ presented here, several other crossings occur, which are not shown in this figure, to avoid 
clumsiness. This point will be further taken up in later part of this section.


\begingroup            
\squeezetable
\begin{table}
\caption{\label{tab:table_crossing}Energy difference and contribution of various components in the pair of states (1s2s $^3$S, 
1s2p$^3$P) and (1s3d$^3$D, 1s2s$^3$S) for confined He atom at different $r_c$'s. See text for details.}
\centering
\begin{ruledtabular}
\begin{tabular}{l|llllllll}
\multicolumn{9}{c}{(1s2p $^3$P, 1s2s $^3$S)}  \\
\hline
  &  $r_c=0.1$ & $r_c=0.3$  & $r_c=0.7$   &   $r_c=1$ &   $r_c=2.5$   &   $r_c=4.4$  &   $r_c=4.5$  & $r_c=20$ \\
\hline
$\Delta E_{(^3\text{P}-^3\text{S})}$   &  $-$941.3233 & $-$99.3402   & $-$16.2298   & $-$7.1798 & $-$0.5294 & $-$0.0004  & 0.0054    & 0.0437     \\
$\Delta T_{(^3\text{P}-^3\text{S})}$   &  $-$964.5660 & $-$107.3300  & $-$19.8379   & $-$9.7766 & $-$1.4554 & $-$0.2810  & $-$0.2603 & $-$0.0425 \\
$\Delta V_{\mathrm{en}_{(^3\text{P}-^3\text{S})}}$  &  25.5989  & 8.7414  & 3.8949   & 2.7744   & 0.9275  & 0.2415  & 0.2265   & 0.0745    \\ 
$\Delta V_{\mathrm{ee}_{(^3\text{P}-^3\text{S})}}$  &   $-$2.3561   & $-$0.7517    & $-$0.2868    & $-$0.1776 & $-$0.0015 & 0.0390     & 0.0391    & 0.0117  \\
\hline 
\multicolumn{9}{c}{(1s3d $^3$D, 1s2s $^3$S)}  \\
\hline
  &  $r_c=0.1$ & $r_c=0.3$  & $r_c=0.5$   &   $r_c=1$ &   $r_c=1.1$   &   $r_c=5$  &   $r_c=10$  & $r_c=30$ \\
\hline
$\Delta E_{(^3\text{D}-^3\text{S})}$   &   $-$284.4894 & $-$63.8781   & $-$6.6476   & $-$0.1066 & 0.1746     &   0.3722    & 0.1637     & 0.1180     \\
$ \Delta T_{(^3\text{D}-^3\text{S})}$   &   $-$313.2606 & $-$78.4721   & $-$12.7192  & $-$3.2871 & $-$2.7324  &   0.0841    & $-$0.0155  & $-$0.1154   \\
$ \Delta V_{\mathrm{en}_{(^3\text{D}-^3\text{S})}}$  &   30.0266     & 15.2141      & 6.3092      & 3.2885    & 3.0031     &   0.3321    & 0.2772     & 0.3948      \\ 
$\Delta V_{\mathrm{ee}_{(^3\text{D}-^3\text{S})}}$  &   $-$1.2554   & $-$0.6200    & $-$0.2376   & $-$0.0959 & $-$0.0440  & $-$0.0259   & $-$0.0980  & $-$0.1613  \\
\end{tabular}
\end{ruledtabular}
\end{table}

\begin{figure}
\centering                       
\begin{minipage}[c]{0.85\textwidth}\centering
\includegraphics[scale=0.85]{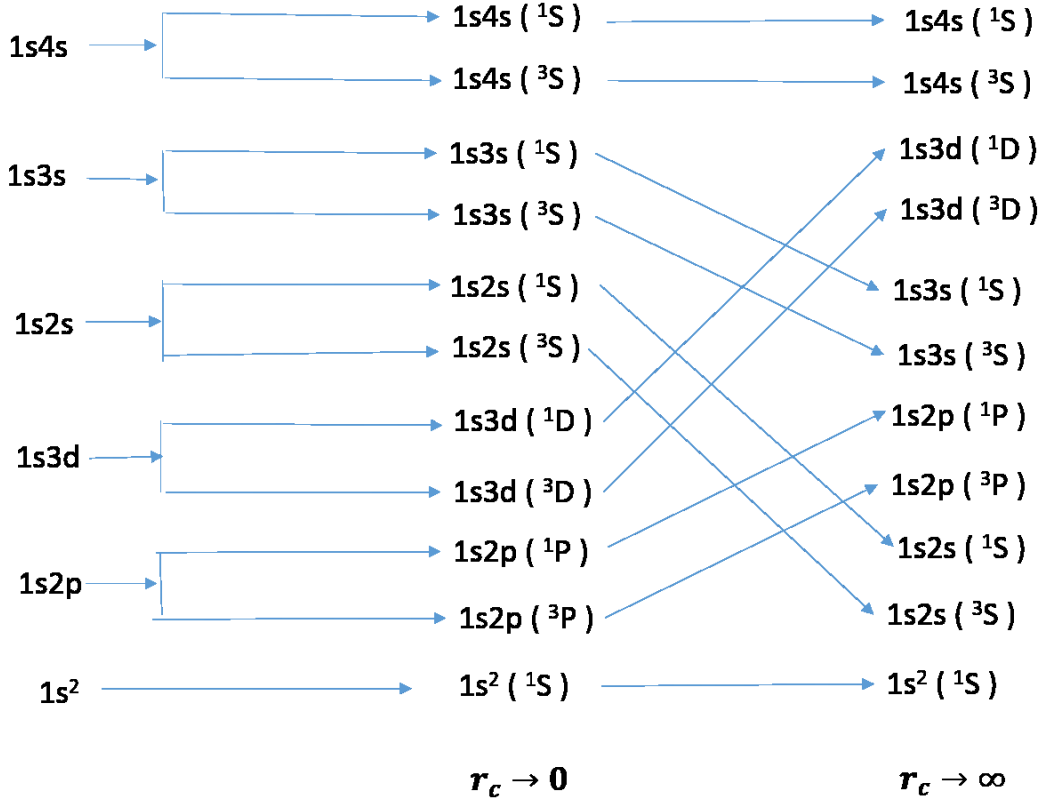}
\end{minipage}%
\caption{\label{fig:figure3}Correlation diagram for a He atom in an impenetrable spherical cavity of radius $r_c$.}
\end{figure}

From the foregoing analysis, it is obvious that energies of a confined He are less sensitive to correlation effect in 
stronger regime; however, with growth in $r_c$ the difference between X-only and correlated energies tends 
to assume greater significance. In Fig.~\ref{fig:figure2}, absolute correlation energy with respect to $r_c$ 
are plotted for some selected states. Panels (a), (b) correspond to Wigner and LYP. In both cases, it is 
noticed that, the nature of correlation energy with compression, maintains similar qualitative pattern for all states, 
for a given functional. In (a), this contribution is found to amplify steadily with confinement strength from free 
atom, becoming more prominent in lower $r_c$. Again, crossing between different states takes place quite frequently at 
several $r_c$'s. The qualitative nature of curves is comparable to the recently published results of 
\cite{sarsa16}. For LYP, however, the plot in (b) considerably alters from (a); as $r_c$ reduces from free 
limit, E$_{\mathrm{corr}}^{\mathrm{LYP}}$ lowers sharply until reaching a minimum, and then it goes up quite 
dramatically at certain $r_c$. Here also few crossovers found, which, again however, alone can not adequately explain 
the observed crossing pattern in energies. 

The interplay between ordering and crossing of states as functions of $r_c$, may be analyzed by constructing a traditional 
correlation diagram, widely used in quantum chemistry \cite{pupyshev17}. Besides the states 
of previous tables, here we also include 1s3s and 1s4s $^{3,1}$S. These are ordered according to their energies
in the limit $r_c \rightarrow 0$ and $r_c \rightarrow \infty$ in middle and right segments. 
This diagram in Fig.~3 consists of three columns, left most of which represents two 
independent particles inside a small sphere. Considering this model to be a starting point, energy ordering for 
states of confined He within a small cavity ($r_c \rightarrow 0$) and free atom ($r_c \rightarrow \infty$) can be 
represented as in second and third columns. In the former case, main contribution to energy is provided by 
kinetic energy, while Coulombic repulsion makes a contribution that is large in absolute value, but 
small in comparison to kinetic energy. The energies and wave functions for a particle in a rigid sphere
are available in standard text book \cite{flugge71}. 
For a system of two independent particles, energy of a state 
having configuration $nl, n'l'$, is given by a sum of orbital energies. 
This simplistic energy-level structure is, however, complicated in presence of an external potential due to a positive 
nucleus of charge $Z$ at the center of cavity. In contrast to the one-electron states of particle inside a sphere, for 
the atomic case, ordering of states does not necessarily remains same at all $r_c$ values. Due to this term in the 
Hamiltonian, triplet and singlet energies separate out, which is shown in mid-section $(r_c \rightarrow 0)$. 
 For the limiting case $r_c \rightarrow \infty$, the energy 
levels are clearly reordered from its confined counterpart. In this scenario, our ordering matches nicely with 
that observed in experiment \cite{sansonetti05}. 
The transition from confinement to free case leads to interesting degeneracy points in correlation diagram as an 
outcome of rearrangement of states. The confinement radius for intersection point cannot be defined from the 
correlation diagram directly. For all states under consideration, the diagram bears qualitative resemblance to that 
reported in \cite{pupyshev17} utilizing a highly accurate, correlated wave function. 

\begin{figure}                         
\centering
\begin{minipage}[c]{0.48\textwidth}\centering
\includegraphics[scale=0.85]{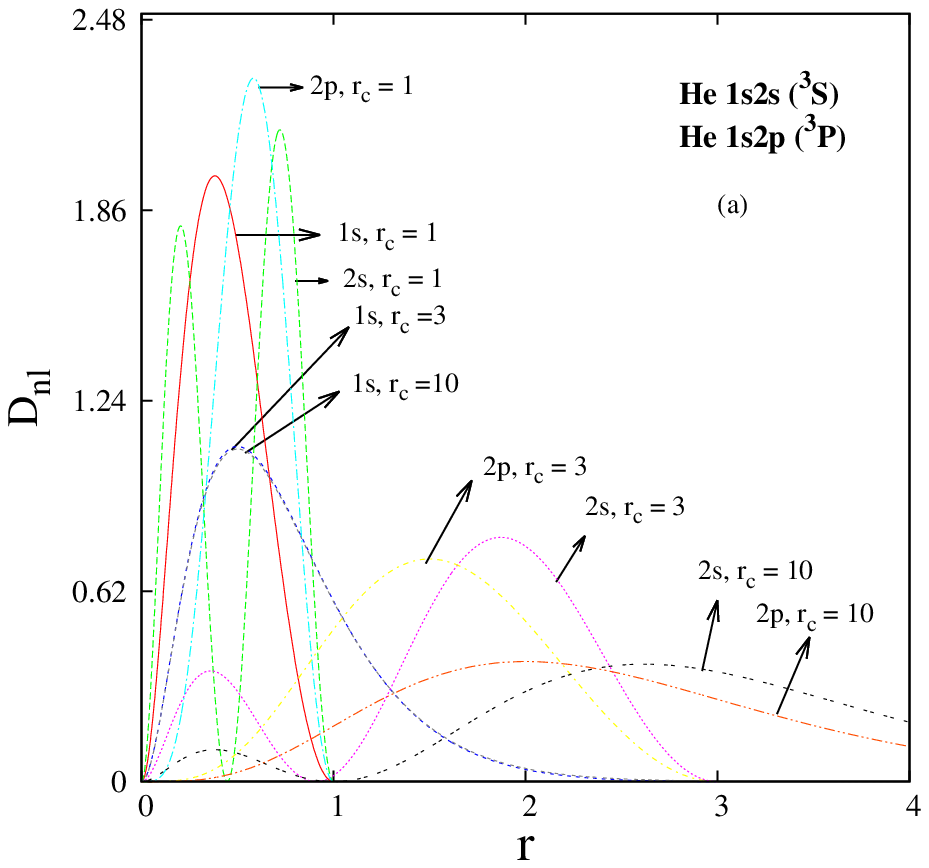}
\end{minipage}%
\begin{minipage}[c]{0.48\textwidth}\centering
\includegraphics[scale=0.85]{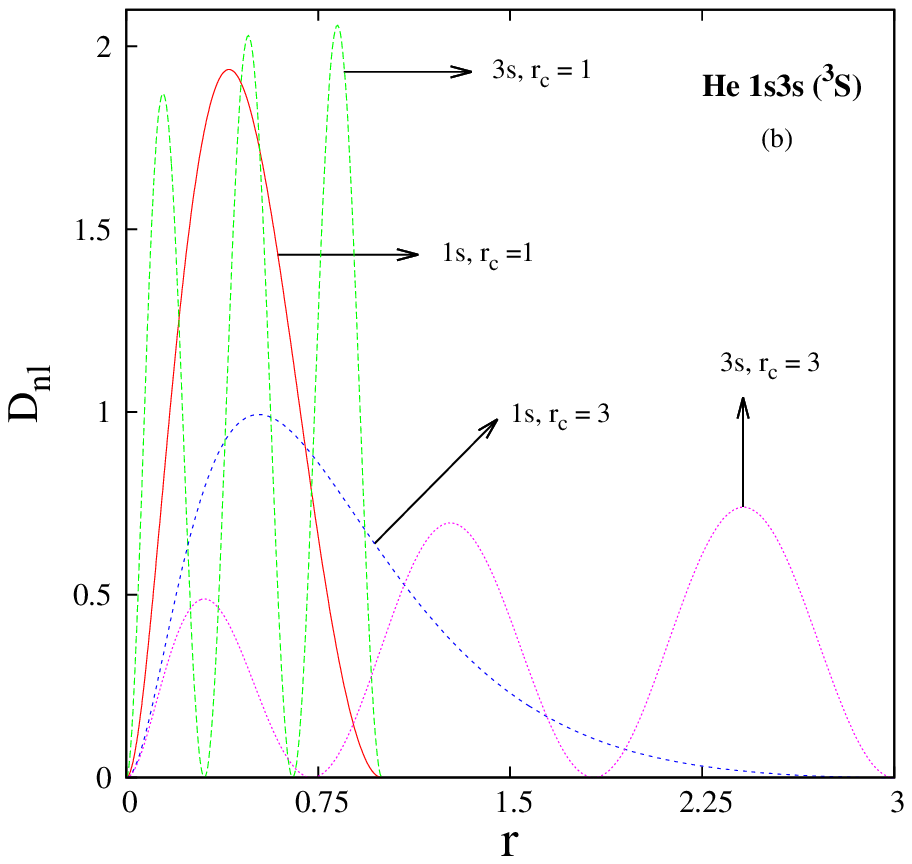}
\end{minipage}%
\caption{\label{fig:figure5}Radial density of orbitals of confined He, at few $r_c$'s in various states: 
(a) 1s2s $^3$S and 1s2p $^3$P (b) \textcolor{red}{1s3s $^3$S }. See text for details.}
\end{figure}

In order to get an estimate of the above crossings among states with changes in $r_c$, in 
Table~\ref{tab:table_crossing}, as an illustration, we present total X-only energy differences, $\Delta E$, between 
(1s2p $^3$P, 1s2s $^3$S) and (1s3d $^3$D, 1s2s $^3$S) pair of states along with various contributions, \emph{viz.}, 
kinetic ($\Delta T$), electron-nucleus ($\Delta V_{\mathrm{en}}$) and electron-electron ($\Delta V_{\mathrm{ee}}$) at
few selected $r_c$. Proceeding from left to right, one approaches stronger confinement to free atom. A change 
of sign occurs in energy difference for ($^3$P, $^3$S) and ($^3$D, $^3$S) in the ranges of $r_c= 4.5$--4.4 and $r_c= 
1.1$--1.0, respectively, indicating a crossover between respective multiplet pairs. For a given state, as $r_c$ 
diminishes, so do both average electron-nucleus and electron-electron distances. This results in a lowering of 
$V_{\mathrm{en}}$, and increase in magnitude of $V_{\mathrm{ee}}$. Also, as a consequence of uncertainty principle, 
as the atom is enclosed in progressively smaller box, $T$ tends to accumulate. The relative contribution of each of 
these terms depends on $r_c$ and the particular state under consideration. The competing effects of these quantities 
lead to the desired crossing between various multiplets of Fig.~\ref{fig:figure3}; major contribution comes 
from one-electron energies $T$ and $V_{\mathrm{en}}$. An analysis of these components helps us conclude that for free 
atom as well as for some sufficiently high $r_c$, $\Delta V_{\mathrm{en}}$ 
is responsible for lower energy of $^3$S than $^3$P and $^3$D. For first pair, it is seen that, at $r_c=20$ or so, 
higher $T$ of $^3$S ($\Delta T < 0$) gets compensated by comparable values of $V_{\mathrm{en}}$ and $V_{\mathrm{ee}}$ 
of $^3$P. In the second pair, for certain $r_c$, $V_{\mathrm{en}}$ also compensates the high $V_{\mathrm{ee}}$; thus 
at $r_c=30$, though $T$ and $V_{\mathrm{ee}}$ both are higher for $^3$S than $^3$D, $V_{\mathrm{en}}$ is responsible
for (+)ve value of $\Delta E$. As we approach stronger confinement, one notices, for these pairs, 
$\Delta V_{\mathrm{ee}}$ hardly contributes to $\Delta E$ relative to $\Delta T$ and $\Delta V_{\mathrm{en}}$. 
A competition between $\Delta T$ and $\Delta V_{\mathrm{en}}$ determines the ordering between terms; $V_{\mathrm{en}}$ 
is higher for $^3$S than both $^3$P, $^3$D, but a rapid increase of $T$ for $^3$S compared to other states, 
leads to the crossing of levels. Another point to be noted is that the data presented in this table is consistent with 
the correlation diagram of Fig.~\ref{fig:figure3}; in free limit, $\Delta E$ between 1s2p $^3$P, 1s2s $^3$S is 
lower than the same between 1s3d $^3$D, 1s2s $^3$S, but due to crossover this pattern changes, and in the 
opposite limit (e.g., $r_c=0.1$), $\Delta E$ of former pair exceeds that of latter. 

\begin{figure}                         
\centering
\begin{minipage}[c]{0.48\textwidth}\centering
\includegraphics[scale=0.75]{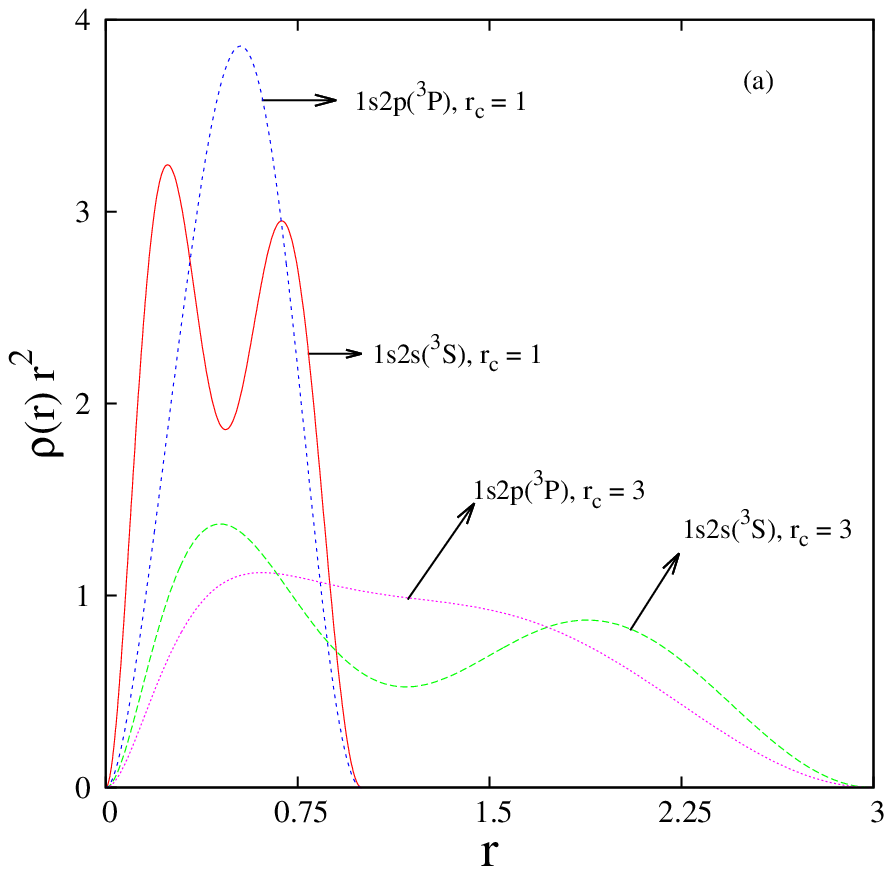}
\end{minipage}%
\begin{minipage}[c]{0.48\textwidth}\centering
\includegraphics[scale=0.75]{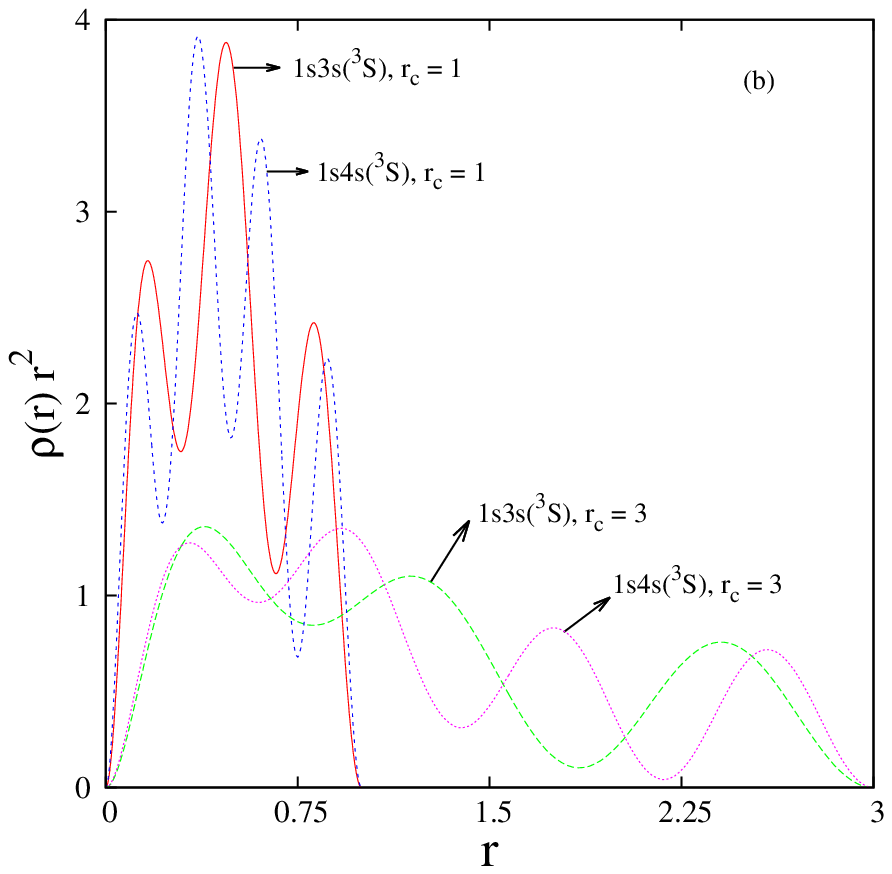}
\end{minipage}%
\caption{\label{fig:figure6}Radial density of (a) 1s2s $^3$S and 1s2p $^3$P (b) 1s3s $^3$S and 1s4s $^3$S 
states of confined He at $r_c=1$ and 3 respectively. Consult text for more details.}
\end{figure}

\begin{figure}
\centering                       
\begin{minipage}[c]{0.32\textwidth}\centering
\includegraphics[scale=0.55]{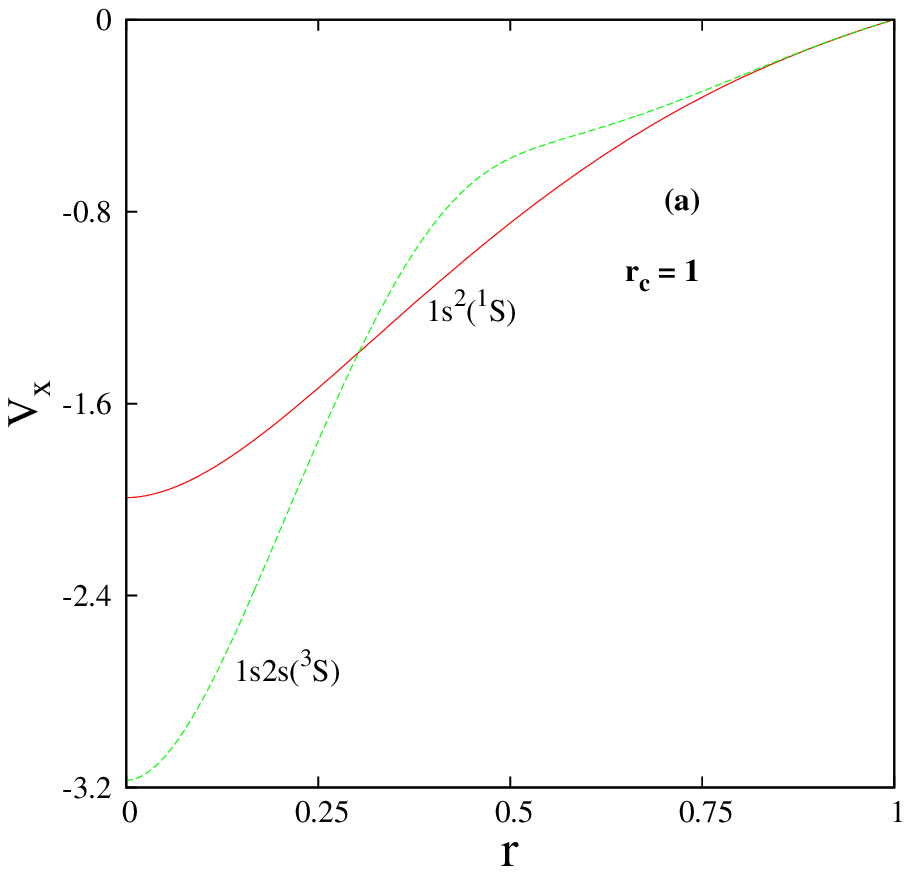}
\end{minipage}%
\begin{minipage}[c]{0.32\textwidth}\centering
\includegraphics[scale=0.55]{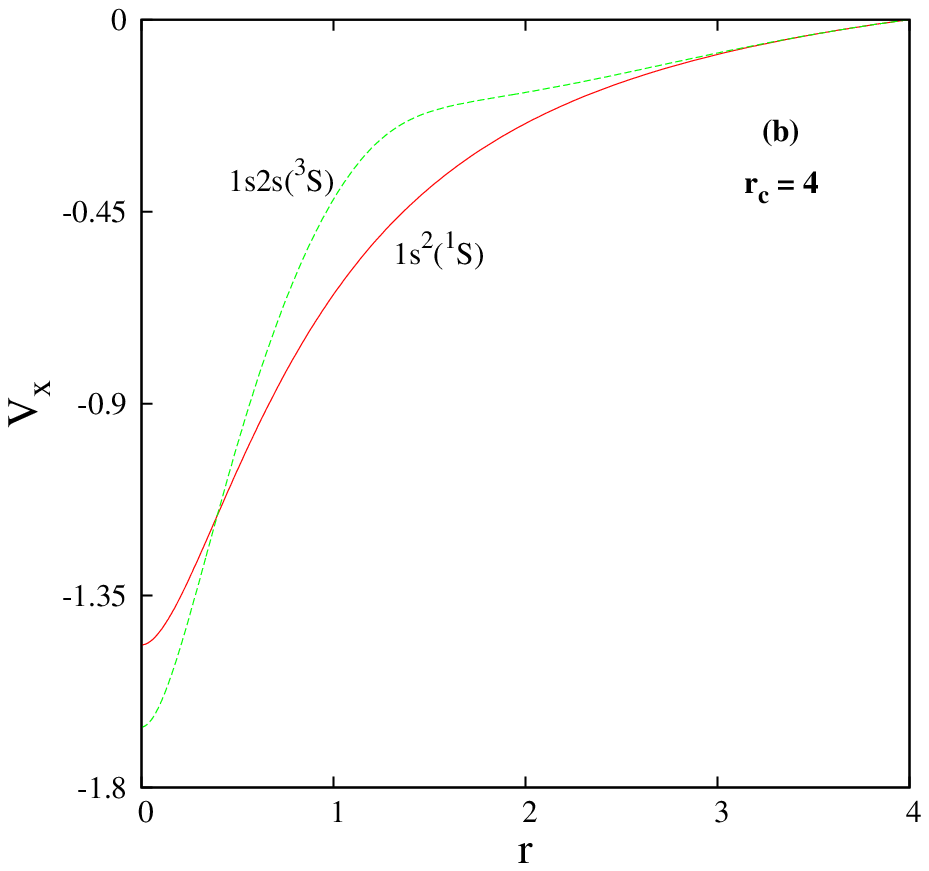}
\end{minipage}%
\begin{minipage}[c]{0.32\textwidth}\centering
\includegraphics[scale=0.55]{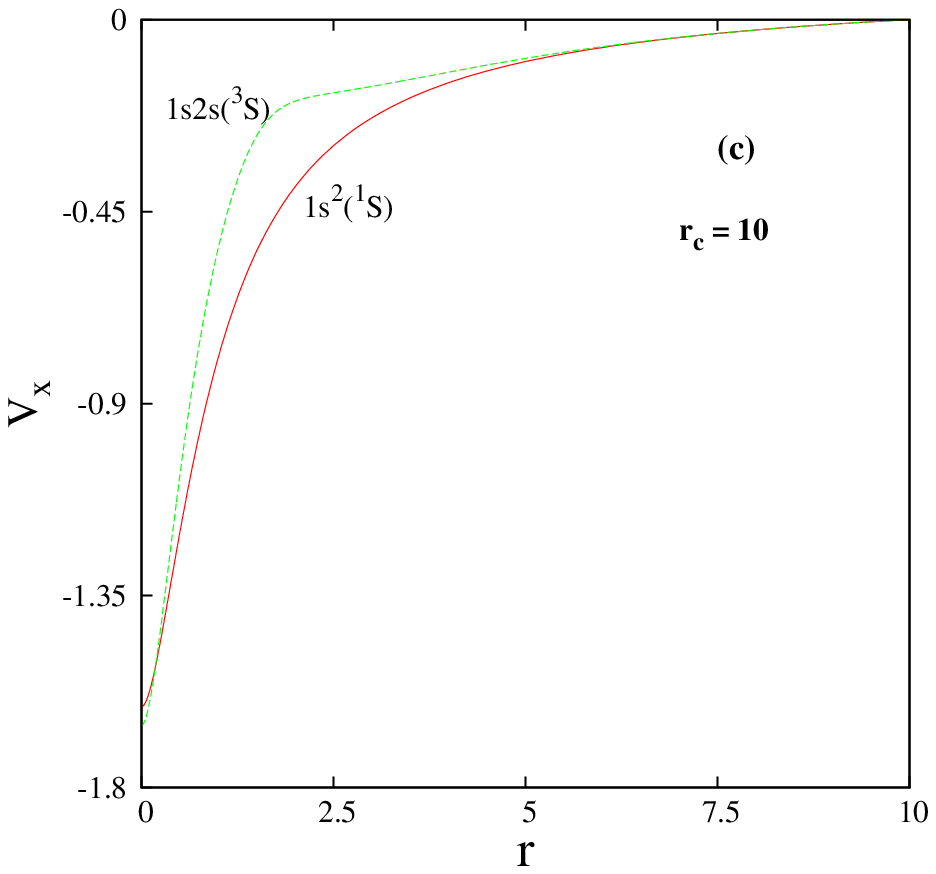}
\end{minipage}%
\caption{\label{fig:figure7}Exchange potential in first singlet and triplet states of confined He at $r_c=1,4,10$. 
in panels (a), (b), (c) respectively. See text for details.}
\end{figure}

Next, in Fig.~\ref{fig:figure5}, we plot the radial probability distribution $D_{nl} = r^2 R_{nl}^2(r)$, for 1s, 2s, 2p 
orbitals associated with 1s2s $^3$S and 1s2p $^3$P states in panel (a), at $r_c$ 1, 3, 10. Similarly (b) gives 
for 1s, 3s orbitals related to 1s3s $^3$S, at $r_c$ 1 and 3. As expected, confinement effects are rather pronounced 
for outermost orbitals in comparison to other orbitals. Thus one notices that 2s, 2p, 3s densities modify significantly
as the atom is squeezed at high pressure. At high $r_c$ (10), area under the first 
peak of 2s is negligible, resulting in radial density of 2s, 2p orbitals being quite similar. As confinement 
strengthens, latter two orbitals begin to differ from each other. Now the area of 2s orbital
assumes significance relative to its free limit, while the maximum of D$_{2p}$(r) lies in between two maxima of 
2s density. For 1s orbital, at $r_c = 3$ and 10, two curves almost overlap, and as we proceed 
to smaller $r_c$ (1), the positions of maxima remain almost unaltered while height of the peaks extend. The 
difference between confined and unconfined charge distributions for 2s, 2p orbitals shows distinct changes. The 
position of nodes and local maxima are shifted to lower $r$'s and magnitudes become considerably larger. This phenomenon 
is more prominent for orbitals in (b) having larger number of nodes than 2s. Thus, in passing from $r_c=3$ 
(weak) to 1 (strong), 1s orbital records a change in peak height, without much shift in its position, while 
for 3s, in going from $r_c = 3$ to 1, significant variations take place in nodal positions, as well as magnitude 
of local maxima. Hence, it is reasonable to assume that, the crossing of energy levels as discussed earlier, may be 
attributed mainly to adjustment of outermost orbitals, since inner orbitals change rather slowly, as
pressure goes up. 

\begingroup            
\squeezetable
\begin{table}
\caption{\label{tab:table4}Density moments (X-only) of 1s$^2$ ($^1$S), 1s2s ($^3$S) states of confined He at 
various $r_c$'s. Numbers in the parentheses denote literature results. See text for details.}
\begin{ruledtabular}

\begin{center}
\begin{tabular}{c|  c c c c c c|c c c c c c}
$r_c$   & \multicolumn{6}{c}{1s$^2$ $^1$S}   &   \multicolumn{6}{c}{1s2s $^3$S} \\
\cline{2-13}
 & $\langle \frac{1}{r^2}\rangle $ & $\langle \frac{1}{r}\rangle $ & $\langle r \rangle $ & $ \langle r^2 \rangle $ & $ \langle r^3 \rangle $ & 
$ \langle r^4 \rangle $ & $\langle \frac{1}{r^2}\rangle$ & $\langle \frac{1}{r}\rangle $ & $\langle r \rangle $ & $ \langle r^2 \rangle $ & 
$ \langle r^3 \rangle $ & $ \langle r^4 \rangle $ \\
\hline
  0.1 & 1866.493  & 49.550 & 0.099 & 0.005 & 0.0003 & 0.00002 & 2865.944 & 56.171 & 0.099 & 0.005 & 0.0004 & 0.00003 \\
  0.3 &  228.700  & 17.099 & 0.290 & 0.048 & 0.008 & 0.001  &  342.610 & 19.183 & 0.295 & 0.053 & 0.010 & 0.002 \\
  0.5 &   91.406  & 10.653 & 0.473 & 0.128 & 0.038 & 0.012  & 133.083  & 11.807 & 0.486 & 0.145 & 0.0184 & 0.017 \\
  0.8 &   42.305  &  7.088 & 0.727 & 0.308 & 0.145 & 0.073  & 58.561   & 7.685  & 0.765 & 0.362 & 0.192 & 0.109 \\
  1.0 &   30.532  &  5.934 & 0.883 & 0.459 & 0.266 & 0.167  & 40.703   & 6.326  & 0.945 & 0.556 & 0.368 & 0.260 \\
	&           &         & (0.883\footnotemark[1]$^,$\footnotemark[2])     & (0.460\footnotemark[1],  & (0.267\footnotemark[1], &  (0.168\footnotemark[1],    &         &        &          &     &   & \\ 
	&           &         &      & 0.459\footnotemark[2])  & 0.267\footnotemark[2]) &  0.167\footnotemark[2])    &         &        &          &     &   & \\ 
  4.0 & 12.081    &  3.393 & 1.829 & 2.271 & 3.523 & 6.484 & 9.510  & 2.645 & 3.024   & 6.494 & 16.232 & 43.767 \\
	&           &         & (1.831\footnotemark[1]$^,$\footnotemark[2])  & (2.272\footnotemark[1]$^,$\footnotemark[2])   & (3.525\footnotemark[1],& 
(6.458\footnotemark[1], &         &        &          &        & &\\  
	&           &         &   &    & 3.524\footnotemark[2]) & 6.456\footnotemark[2]) &         &        &          &        & &\\     
  5.0 & 12.025    &  3.380 & 1.848 & 2.346 & 3.785 & 7.393 & 8.913  & 2.501 & 3.498   & 9.069 & 27.639 & 91.174 \\
	&           &         & (1.850\footnotemark[1]$^,$\footnotemark[2])  & (2.349\footnotemark[1]$^,$\footnotemark[2]) , & (3.790\footnotemark[1],& 
(7.371\footnotemark[1], & & & & & &      \\
	&           &         &   &  & 3.789\footnotemark[2]) & 7.366\footnotemark[2]) & & & & & &      \\
  8.0 & 12.019    &  3.378 & 1.852 & 2.363 & 3.858 & 7.717 & 8.442  & 2.353 & 4.472   & 16.324 & 71.987 & 349.170 \\
	&           &         & (1.855\footnotemark[1]$^,$\footnotemark[2])  & (2.370\footnotemark[1], & (3.880\footnotemark[1], &  (7.765\footnotemark[1],       &         &        &          &     &   & \\
	&           &         &  & 2.369\footnotemark[2]) & 3.878\footnotemark[2]) &  7.756\footnotemark[2])       &         &        &          &     &   & \\
	$\infty$ & 12.0 & 3.37\footnotemark[3] & 1.851\footnotemark[3] & 2.362\footnotemark[3]  & 3.85 & 7.70 & 8.365 & 2.318 & 4.994 & 21.914 & 122.04 & 786.72 \\
	 &        &         & (1.855\footnotemark[1]$^,$\footnotemark[2])  & (2.373\footnotemark[1],  & (3.891\footnotemark[1], & (7.803\footnotemark[1], &  & &  & & & \\
	 &        &         &   & 2.372\footnotemark[2])  & 3.886\footnotemark[2]) & 7.777\footnotemark[2]) &  & &  & & & \\

\end{tabular}
\end{center}
\end{ruledtabular}
\begin{tabbing}
$^{\mathrm{a}}$HF result \cite{vyboishchikov15}. \hspace{25pt}  \=  
$^{\mathrm{b}}$ZMP result \cite{vyboishchikov15}. \hspace{25pt}  \= 
$^{\mathrm{c}}$The corresponding HF result \cite{fischer77}, with density normalized to unity, for \\
$\langle \frac{1}{r} \rangle$, $\langle r \rangle$ and $\langle r^2 \rangle$ are 1.1544, 2.5599 and 11.5612. Present 
estimates are: 1.1578, 2.5030, 11.0112. 
\end{tabbing}
\end{table}
\endgroup 

The effect of pressure on radial densities is nicely depicted in Fig.~\ref{fig:figure6}: panel (a) shows this 
for 1s2p $^3$P and 1s2s $^3$S, while (b) for 1s3s $^3$S and 1s4s $^3$S, respectively, at two representative $r_c$, 
namely, 1 and 3. One notices 4, 3, 2 and 1 maxima in 1s4s, 1s3s, 1s2s and 1s2p states. For a given state, with 
increasing pressure, the positions of these maxima get shifted to lower $r$, peaks become narrower and enhance in 
magnitude.  

To pursue further, it is worthwhile to study the performance of present exchange potential.
It has been compared with other accurate potentials in literature, which reproduce HF 
electron density. Some such examples are BJ model potential, ZMP, B88 and SC$\alpha$-LDA 
method. As such, $v_x(r)$ is calculated at each point within the work-function approximation, by solving the KS equation 
self-consistently for a given $r_c$. In Fig.~\ref{fig:figure7}, it is displayed at three $r_c$ values (1, 4, $\infty$), 
for ground and 1s2s ($^3$S) states of He. Its behavior in confined environment is visibly different from that in the 
free limit; with lowering of $r_c$ it becomes steeper. For $^3$S state there is a distinct hump in 
the curve, which gets flatter with a reduction in confinement strength. An analysis of \cite{vyboishchikov15} reveals
that, BJ and ZMP potentials perform quite well in reproducing the exact exchange for a two-electron confined system 
in ground state. A comparison of our figure confirms that $v_x (r)$ at $r_c=4$ qualitatively reproduces the 
characteristic features found in the literature \cite{vyboishchikov15}.  

\begingroup          
\squeezetable
\begin{table}
\caption{\label{tab:table5}Ground-states energies of Li$+$ and Be$^{2+}$ in spherical cavity of various $r_c$'s.}
\begin{ruledtabular} 
\begin{tabular}{c|c| c c c|c| c c c}
& \multicolumn{4}{c|}{Li$^{+}$} & \multicolumn{4}{c}{Be$^{2+}$} \\
\hline
$r_c$  & X-only       & XC-Wigner   & XC-LYP      & Literature  &  X-only      & XC-Wigner        & XC-LYP  & Literature \\
\hline
0.5    & 11.8249      & 11.7234     & 11.9715     & 11.7790\footnotemark[2],11.7768\footnotemark[3],      
       & 0.1525       & 0.0492      & 0.2767      & 0.1078\footnotemark[2],0.1056\footnotemark[3],\\
       & 11.82550\footnotemark[1]   &             &        &   11.76771\footnotemark[4] 
       & 0.15261\footnotemark[1]             &             &             & 0.10801980\footnotemark[4] \\ 
0.6    & 3.9733       & 3.8798      &  4.0774     & 3.9284\footnotemark[2],3.9262\footnotemark[3],      
       & $-$6.1964    & $-$6.2923   & $-$6.1153   & $-$6.2402\footnotemark[2],$-$6.2423\footnotemark[3],\\
       & 3.97368\footnotemark[1]              &             &             & 3.9934290\footnotemark[4] 
       & $-$6.19641\footnotemark[1]           &             &             & $-$6.288159\footnotemark[4],$-$6.242352\footnotemark[5]   \\ 
0.7    & $-$0.3149    & $-$0.4019   & $-$0.2422   & $-$0.3590\footnotemark[2],$-$0.3611\footnotemark[3]
       & $-$9.4540    & $-$9.5441   & $-$9.4046   & $-$9.4970\footnotemark[2],$-$9.4991\footnotemark[3], \\
       & $-$0.31480\footnotemark[1]              &             &             & $-$0.361136\footnotemark[5] 
       & $-$9.45401\footnotemark[1]              &             &             & $-$9.499124\footnotemark[5]\\
0.8    & $-$2.8178    & $-$2.8996   & $-$2.7692   & $-$2.8612\footnotemark[2],$-$2.8632\footnotemark[3],       & $-$11.2234   & $-$11.3091  & $-$11.1979  & $-$11.2658\footnotemark[2],$-$11.2679\footnotemark[3], \\
       &              &             &             & $-$2.893898\footnotemark[4],$-$2.863228\footnotemark[5]  
       &              &             &             & $-$11.26839\footnotemark[4],$-$11.267912\footnotemark[5]   \\ 
0.9    & $-$4.3490    & $-$4.4265   & $-$4.3192   & $-$4.3918\footnotemark[2],$-$4.3937\footnotemark[3],
       & $-$12.2204   & $-$12.3027  & $-$12.2131  & $-$12.2624\footnotemark[2],$-$11.2645\footnotemark[3]                    \\ 
       & $-$4.34907\footnotemark[1]              &             &             & $-$4.393732\footnotemark[5] 
       & $-$12.22028\footnotemark[1]             &             &             &              \\
1.0    & $-$5.3183    & $-$5.3922   & $-$5.3033   & $-$5.3605\footnotemark[2],$-$5.3635\footnotemark[7],
       & $-$12.7954   & $-$12.8750  & $-$12.8021  & $-$12.8369\footnotemark[2], $-$12.8393\footnotemark[3],     \\
       & $-$5.31832\footnotemark[1]&             & & $-$5.3624\footnotemark[3],$-$5.362399\footnotemark[5]$^,\footnotemark[6]$
       & $-$12.79481\footnotemark[1]&             &             & $-$12.839307\footnotemark[5]     \\
       & $-$5.318324\footnotemark[6] &             &             & 
       &              &             &             &              \\
1.2    & $-$6.3632    & $-$6.4318   & $-$6.3698   & $-$6.4047\footnotemark[2], $-$6.4065\footnotemark[3],       & $-$13.3295   & $-$13.4057  & $-$13.3550  & $-$13.3701\footnotemark[2], $-$13.3733\footnotemark[3], \\
       & $-$6.36317\footnotemark[1] &             &             & $-$6.407358\footnotemark[4]       & $-$13.32825\footnotemark[1]&             &             & $-$13.36396\footnotemark[4]   \\ 
1.4    & $-$6.8302    & $-$6.8953   & $-$6.8509   & $-$6.8713\footnotemark[2], $-$6.8732\footnotemark[3],       & $-$13.5151   & $-$13.5894  & $-$13.5515  & $-$13.5552\footnotemark[2], $-$13.5590\footnotemark[3], \\
       & $-$6.82981\footnotemark[1]  &             &             &  $-$6.855290\footnotemark[4] 
       & $-$13.51486\footnotemark[1] &             &             & $-$13.556580\footnotemark[4]   \\ 
1.6    & $-$7.0462                   & $-$7.1090   & $-$7.0761   & $-$7.0869\footnotemark[2],$-$7.0892\footnotemark[3] 
       & $-$13.5791                  & $-$13.6526  & $-$13.6215  & $-$13.6193\footnotemark[2],$-$13.6232\footnotemark[3]          \\ 
       & $-$7.04585\footnotemark[1]              &             &             & 
       & $-$13.57911\footnotemark[1]             &             &             &              \\
1.8    & $-$7.1475    & $-$7.2089   & $-$7.1834   & $-$7.1880\footnotemark[2], $-$7.1906\footnotemark[3],       & $-$13.6007   & $-$13.6738  & $-$13.6465  & $-$13.6415\footnotemark[2], $-$13.6449\footnotemark[3], \\
       & $-$7.14643\footnotemark[1]              &             &             & $-$7.192726\footnotemark[4]      & $-$13.60028\footnotemark[1]             &             &             & $-$13.639290\footnotemark[4]   \\ 
2.0    & $-$7.1952    & $-$7.2556   & $-$7.2348   & $-$7.2356\footnotemark[2]$^,\footnotemark[7]$,$-$7.2383\footnotemark[3],    
       & $-$13.6079   & $-$13.6808  & $-$13.6555  & $-$13.6493\footnotemark[2],$-$13.6521\footnotemark[3]   \\
       & $-$7.19460\footnotemark[1]              &             &             & $-$7.238402\footnotemark[5] 
       & $-$13.61002\footnotemark[1]             &             &             &              \\
2.5    & $-$7.2306    & $-$7.2901   & $-$7.2750   & $-$7.2740\footnotemark[3],$-$7.258839\footnotemark[4]
       & $-$13.6110   & $-$13.6838  & $-$13.6611  & $-$13.6553\footnotemark[3],$-$13.654130\footnotemark[4]   \\
3.5    & $-$7.2362    & $-$7.2955   & $-$7.2836   & $-$7.2791\footnotemark[3],$-$7.279050\footnotemark[4] 
       & $-$13.6112   & $-$13.6840  & $-$13.6633  & $-$13.6555\footnotemark[3],$-$13.655678\footnotemark[4]   \\
5.0    & $-$7.2363    & $-$7.2956   & $-$7.2852   & $-$7.2784\footnotemark[2],$-$7.2783\footnotemark[7],       & $-$13.6112   & $-$13.6840  & $-$13.6644  & $-$13.6539\footnotemark[2], $-$13.6555\footnotemark[3],\\
       & $-$7.236415\footnotemark[6]&             & & $-$7.2798\footnotemark[3],$-$7.279254\footnotemark[4],   
       &              &             &             & $-$13.657630\footnotemark[4],$-$13.655566\footnotemark[5]      \\
	&  &    &    &   $-$7.279913\footnotemark[5]$^,\footnotemark[6]$  &    &    &     &      \\ 
7.0    & $-$7.2363    & $-$7.2956   & $-$7.2860   & $-$7.2784\footnotemark[2],$-$7.279913\footnotemark[5]  
       & $-$13.6112   & $-$13.6840  & $-$13.6651  & $-$13.6539\footnotemark[2],$-$13.655566\footnotemark[5] \\
$\infty$ & $-$7.2363  & $-$7.2956   & $-$7.2876   & $-$7.2799\footnotemark[7]$^,\footnotemark[3]$,$-$7.279913\footnotemark[5]$^,\footnotemark[6]$ 
       &  $-$13.6112  & $-$13.6840  & $-$13.6665  & $-$13.6555\footnotemark[3],$-$13.655566\footnotemark[5]  \\
       & $-$7.236415\footnotemark[6]&             & & 
       &              &             &             &      \\
\end{tabular}
\end{ruledtabular}
\begin{tabbing}
$^{\mathrm{a}}$Ref.~\cite{yakar11}. \hspace{25pt}  \=  
$^{\mathrm{b}}$Ref.~\cite{ludena79}. \hspace{25pt}  \= 
$^{\mathrm{c}}$Ref.~\cite{flores-riveros08}. \hspace{25pt}  \=  
$^{\mathrm{d}}$Ref.~\cite{doma12} \hspace{25pt}  \= 
$^{\mathrm{e}}$Ref.~\cite{bhattacharyya13}. \hspace{25pt}  \=  
$^{\mathrm{f}}$Ref.~\cite{wilson10}. \hspace{25pt}  \= 
$^{\mathrm{g}}$Ref.~\cite{joslin92}.   
\end{tabbing}
\end{table}
\endgroup  

Simply looking at $\rho(r)$ plots does not provide a complete picture, as possible deviation between different
densities can be sometimes very small compared with modulations in a given density at various $r$. For this, one 
resorts to some of the indicators (in the form of density moments) which quantitatively characterize the density 
distribution in a compact manner, and offer valuable insights. To this end, six expectation 
values, $\langle \frac{1}{r^2} \rangle$, $\langle \frac{1}{r} \rangle$, $\langle r \rangle$, 
$\langle r^2 \rangle$, $\langle r^3 \rangle$, $\langle r^4 \rangle$ are offered for ground (left) and 1s2s $^3$S (right) 
states of boxed-in He, at nine selected $r_c$'s within the X-only framework. Only some scattered results are available 
for $\langle r \rangle$, $\langle r^2 \rangle$, $\langle r^3 \rangle$, $\langle r^4 \rangle$ in ground state, where 
the obtained moments match excellently with those of ZMP and HF values \cite{vyboishchikov15}; in some occasions, 
they are completely identical. In free atom ground state, numerical 
HF results \cite{fischer77} for $\langle \frac{1}{r} \rangle$, $\langle r \rangle$, $\langle r^2 \rangle$ compare very 
nicely with current work. For excited state, no published results could be found, to the best of our 
knowledge. Hence, the proposed approach offers accurate moments in both confined and free atom.

\subsection{Confined He-isoelectronic series}
Now, we aim to analyze the role played by nuclear charge, $Z$ on the properties of atom under varying confinement 
strengths. In this regard, in Table~\ref{tab:table5}, ground-state energies, without and with effects of 
correlation, are recorded for two members of He iso-electronic series (Li$^+$, Be$^{2+}$) at representative $r_c$, 
along with reference theoretical results, wherever feasible. As expected, for all systems, stronger confinement 
leads to enhanced total energies. The X-only results of columns 2 and 6 are compared with HF calculation \cite{yakar11} 
throughout the entire $r_c$; also for $Z=3$, at $r_c=1, 5$ as well as the free atom, with that of \cite{wilson10}. 
The agreement with both these references is extremely good. The correlated values of two ionic species, on the 
other hand, can be compared with CI \cite{ludena79} and various Monte-Carlo \cite{joslin92,doma12} methods, in 
addition to Hylleraas-type \cite{flores-riveros08, wilson10, bhattacharyya13} results. The general trend is similar 
to that found for He. At larger $r_c$, difference between Wigner and LYP remains rather quite small, where both 
slightly underestimate literature energies. In stronger confinement region, Wigner energies appears to 
have a marginal edge over LYP. Taking the variational Monte-Carlo \cite{doma12} as reference, discrepancies appear 
to be slightly higher in larger $r_c$. As $Z$ goes up, electron density becomes more contracted; this can be looked 
as a shrinkage of atomic dimensions due to presence of a (+)ve charge at nucleus, which is akin to the compression of 
atom. However, the difference between this process and the one created by 
putting it inside an impenetrable cavity of varying radius is that, by enhancing $Z$, the pressure is induced 
centrally, whereas in latter case it is exerted peripherally. This difference is observed in energy pattern as 
well; the induced pressure due to growing $Z$ eventually lowers the total energy at a fixed confinement strength.  

\begin{figure}
\centering                       
\vspace{1.2cm}
\begin{minipage}[t]{0.48\textwidth}\centering
\includegraphics[scale=0.75]{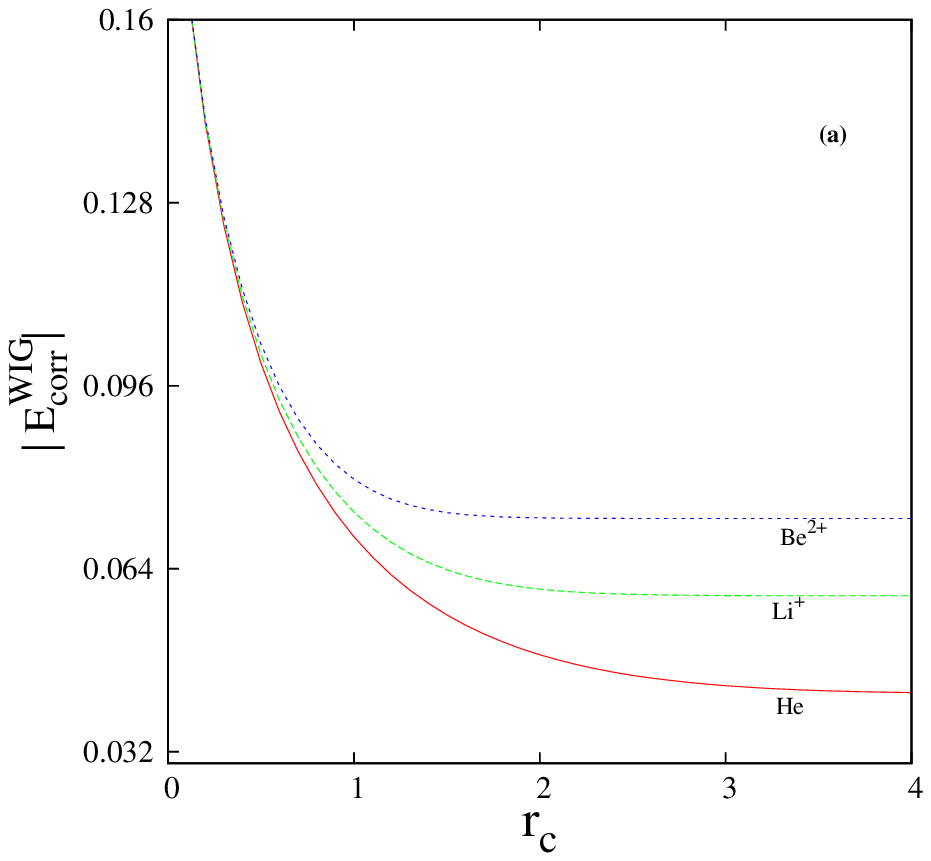}
\end{minipage}
\begin{minipage}[t]{0.48\textwidth}\centering
\includegraphics[scale=0.75]{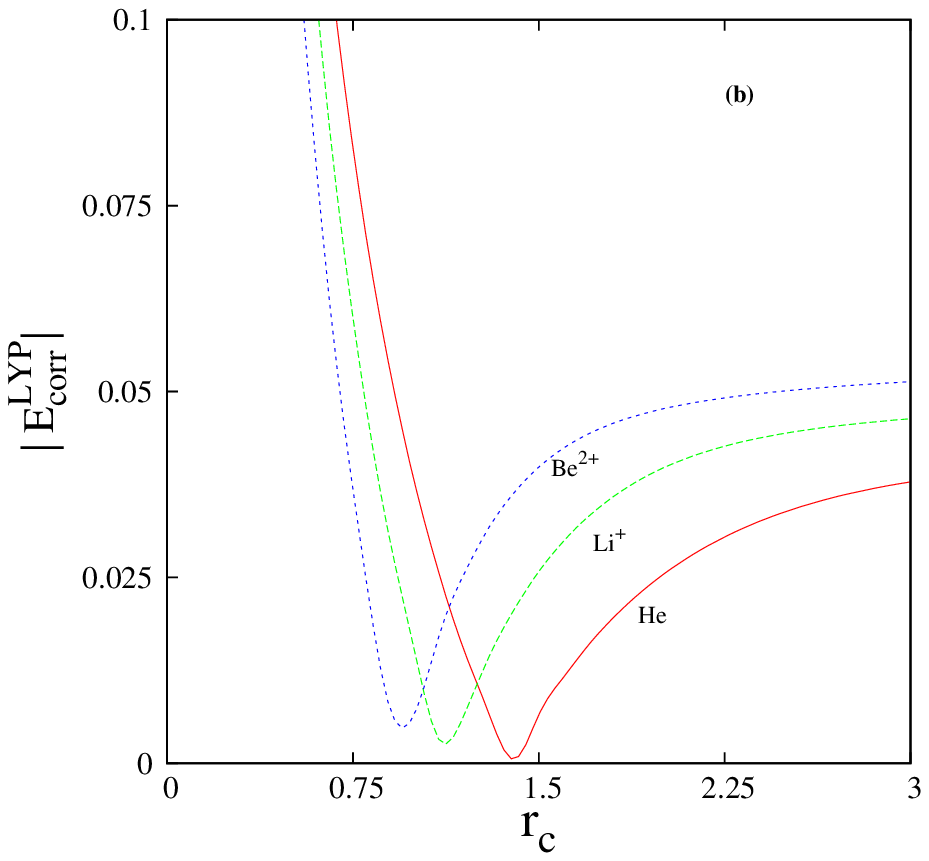}
\end{minipage}
\caption{\label{fig:figure8}Absolute correlation energy, with Wigner (a) and LYP (b) functionals, in the lowest 
states of He, Li$^+$ and Be$^{2+}$. See text for details.}
\end{figure} 

\begingroup            
\squeezetable
\begin{table}
\caption {\label{tab:table7}Ground-state energies of Li and Be, confined at the center of an impenetrable spherical cavity of 
radius $r_c$. See text for details.}
\begin{ruledtabular} 
\begin{tabular}{c|c c | c c c|c c| c c }
& \multicolumn{5}{c|}{Li} & \multicolumn{4}{c}{Be} \\
\cline{2-10}
$r_c$ & $X-$only   & Literature      & XC-Wigner      & XC-LYP    & Literature &  X-only    & Literature    & XC-Wigner     & XC-LYP   \\
\hline
0.5   & 78.4392    &                           & 78.2803   & 78.7693   &            & 124.2082   &                         & 123.9863      & 124.6458 \\
0.6   & 48.4997    &                           & 48.3534   & 48.7539   &            & 75.1709    &                         & 74.9657       & 75.5088 \\
0.7   & 31.1262    &                           & 30.9903   & 31.3236   &            & 46.8218    &                         & 46.6305       & 47.0847 \\
0.8   & 20.2839    &                           & 20.1567   &20.4372    &            & 29.1990    &                         & 29.0193       & 29.4035 \\
1.0   & 8.2385     &$-$8.5139\footnotemark[1]  &  8.1249   & 8.3285    &            &  9.7334    & 9.7327\footnotemark[2], &  9.5720       &  9.8537 \\
      &            &                           &           &           &            &            & 9.8351\footnotemark[5]  &               &         \\
1.2   & 2.2191     & 2.2191\footnotemark[4]    &  2.1155   & 2.2663    &            & 0.0917     & 0.1368\footnotemark[5]  & $-$0.0560     & 0.1550   \\
1.5   & $-$2.2278  &$-$2.2281\footnotemark[2], &$-$2.3208  &$-$2.2221  &$-$1.9085\footnotemark[3],  &$-$6.9464   & $-$6.9477\footnotemark[2],&$-$7.0796           &$-$6.9391 \\
      &            &                           &           &           &$-$1.9870\footnotemark[7]   &            &  $-$6.92237\footnotemark[5] & & \\
2.0   & $-$5.1780  &$-$5.1782\footnotemark[2], &$-$5.2606  &$-$5.2099  & $-$5.1305\footnotemark[3], &$-$11.5064  & $-$11.5079\footnotemark[2], &$-$11.6244          &$-$11.5503 \\
      &            & $-$5.1780\footnotemark[4],&           &           & $-$5.1251\footnotemark[7]  &            & $-$11.4895\footnotemark[5] & & \\   
      &            &$-$5.0841\footnotemark[1]  &           &           &                             &            & $-$11.5078\footnotemark[6] & & \\           
2.5   & $-$6.2951  & $-$6.2955\footnotemark[4] &$-$6.3721  &$-$6.3451  &                             &$-$13.1567  & $-$13.1583\footnotemark[2],  &$-$13.2660           &$-$13.2261 \\
      &            &                           &           &           &                             &             & $-$13.1583\footnotemark[6], & &  \\ 
      &            &                           &           &           &                             &             & $-$13.1413\footnotemark[5]  & &   \\ 
3.0   & $-$6.8018  &$-$6.8027\footnotemark[2]  &$-$6.8753  &$-$6.8609  &$-$6.8304\footnotemark[3],   &$-$13.8614   & $-$13.8613\footnotemark[2], &$-$13.9651          &$-$13.9441 \\
      &            &$-$6.7463\footnotemark[1], &           &           &$-$6.8297\footnotemark[7]    &             & $-$13.8631\footnotemark[6], & & \\ 
      &            &$-$6.8026\footnotemark[4]  &           &           &                             &             &  $-$13.8468\footnotemark[5] &  & \\
4.0   & $-$7.2035  &$-$7.2046\footnotemark[2], &$-$7.2731  &$-$7.2694  &$-$7.2448\footnotemark[3],   &$-$14.3670   & $-$14.3685\footnotemark[2],  &$-$14.4645          &$-$14.4606 \\
      &            &$-$7.1859\footnotemark[1], &           &           &$-$7.2442\footnotemark[7]    &             & $-$14.3678\footnotemark[6],& &  \\ 
      &            &                           &           &           &                             &            &$-$14.3521\footnotemark[5] & & \\ 
5.0   & $-$7.3383  &$-$7.3395\footnotemark[2], &$-$7.4060  &$-$7.4058  &$-$7.3815\footnotemark[3],   &$-$14.5080  & $-$14.5091\footnotemark[2],  &$-$14.6024          &$-$14.6046 \\
      &            &$-$7.3230\footnotemark[1], &           &           &$-$7.3815\footnotemark[7]    &            & $-$14.4918\footnotemark[5] &  &  \\ 
      &            &$-$7.2045\footnotemark[4]  &           &           &                             &            &                            &  &   \\   
6.0   & $-$7.3913  &$-$7.3925\footnotemark[2]  &$-$7.4577  &$-$7.4588  &$-$7.4342\footnotemark[3],   &$-$14.5515  & $-$14.5522\footnotemark[2], &$-$14.6443          &$-$14.6488 \\
      &            &$-$7.3769\footnotemark[1]  &           &           &$-$7.4343\footnotemark[7]    &            &  $-$14.5345\footnotemark[5]  & &   \\
8.0   & $-$7.4238  &$-$7.4249\footnotemark[2], &$-$7.4890  &$-$7.4903  &$-$7.4658\footnotemark[3],   &$-$14.5695  & $-$14.5535\footnotemark[5],  &$-$14.6612          &$-$14.6672 \\
      &            &$-$7.4098\footnotemark[1], &           &           &$-$7.4657\footnotemark[7]    &            & $-$14.5704\footnotemark[2]  &  &   \\ 
      &            &$-$7.4246\footnotemark[4]  &           &           &                             &            &                             &  &   \\   
10.0  & $-$7.4301  &$-$7.4717\footnotemark[2], &$-$7.4948  &$-$7.4962  &                             &$-$14.5711  & $-$14.5729\footnotemark[6],  &$-$14.6626          & $-$14.6692 \\ 
      &            &$-$7.4165\footnotemark[1]  &           &           &                             &            & $-$14.5567\footnotemark[5]  &  & \\     
\end{tabular}
\end{ruledtabular}
\begin{tabbing}
$^{\mathrm{a}}$Ref.~\cite{sanu-ginarte19} \hspace{25pt}  \=
$^{\mathrm{b}}$Ref.~\cite{ludena78} \hspace{25pt}  \=  
$^{\mathrm{c}}$Ref.~\cite{sarsa11} \hspace{25pt}  \= 
$^{\mathrm{d}}$Ref.~\cite{sarsa14} \hspace{25pt}  \= 
$^{\mathrm{e}}$Ref.~\cite{sanu-ginarte18}   \hspace{25pt} \=
$^{\mathrm{f}}$Ref.~\cite{rodriguez15}  \hspace{25pt}  \= 
$^{\mathrm{g}}$Ref.~\cite{lesech11}
\end{tabbing}
\end{table}
\endgroup  

A few words may now be devoted to the influence of $Z$ on correlation energy as confinement takes
place. Thus Fig.~\ref{fig:figure8} depicts how $|$E$_{\mathrm{corr}}|$, approximated by Wigner and LYP functional,
behaves with variation of $r_c$, in case of ground states of He, Li$^+$ and Be$^{2+}$, in panels (a) and (b). 
Starting from the limiting case of free ion, E$_{\mathrm{corr}}^{\mathrm{WIG}}$, for a given member
of iso-electronic series, remains practically unaffected until about $r_c \approx 2$ a.u.; any further 
compression is accompanied by a sharp increase. As $Z$ passes from 2 to 4, with lowering in $r_c$, the 
differences in $E_{\mathrm{corr}}^{\mathrm{WIG}}$ between two successive members reduce. As in stronger 
confinement regime, both nucleus-electron and electron-electron distances fall down; hence the effect of $Z$ gets 
dominated by the confining potential. This results in the fact that as $r_c$ declines, the plots very nearly 
overlap with each other. These observations reinforce the inferences drawn in our recent report \cite{majumdar20}. In 
contrast, however, panel (b) shows a distinct minimum in E$_{\mathrm{corr}}^{\mathrm{LYP}}$ vs $r_c$ graph for 
all the species. As the cavity gets smaller, its magnitude at first diminishes steadily, then passes through a 
minimum and ultimately rises abruptly. This minimum is deeper and wider for He than Li$^+$, which in turn, has 
greater measure than Be$^{2+}$. The position of this minimum moves to lower $r_c$, as $Z$ advances.    


\subsection{Confined Li and Be atoms}
Finally, we use our method to study more than two-electron systems; as sample cases we present the results for Li and Be atoms 
confined in an impenetrable sphere. The results for this confined three- and four-electron systems respectively, are reported for 
both X-only and correlated cases separately in Table~\ref{tab:table7} for ground state. For sake of comparison, along with our 
X-only results of Li in column 2, we provide the results from HF \cite{ludena78}, direct variational \cite{sanu-ginarte19} 
and POEP methods \cite{sarsa14} in column 3. As it is visible from the table, reference results are more prevalent in the 
region $r_c \geq 1$ than $r_c \leq 1$. The presented data for our X-only calculation are in excellent agreement with those 
of \cite{ludena78}. Comparison with \cite{sanu-ginarte19} and \cite{sarsa14} also show good matching.
The correlated energies can be compared with variational Monte Carlo method \cite{sarsa11} and Rayleigh-Ritz variational 
approach \cite{lesech11}. For both functionals, with decreasing value of $r_c$ the difference between present and 
reference energy tends to accumulate. Furthermore, at smaller $r_c$ region, Wigner and LYP differ from each other  
significantly. Results reported for X-only energy values for Be atom are also in very good harmony with HF results 
of \cite{ludena78, rodriguez15}. No correlated results could be found in literature for confined Be.  
Also all the mentioned references are of wave function based method; no DFT calculation of these systems are found. 

\section{Concluding Remarks}
A simple general and accurate KS DFT method has been proposed for calculation of an atom enclosed inside
a rigid spherical cavity of varying radius. The prescription is computationally feasible and can be easily 
extended to other atoms/states. Properties such as energy, radial density, 
expectation values are reported for ground and singly excited states (1s2s $^{3,1}$S, 1s2p $^{3,1}$P, 1s3d $^{3,1}$D) 
of He, as well as ground states of Li$^+$, Be$^{2+}$, in weak, intermediate, strong confinement strengths.
Moreover, ground state results are also reported for Li and Be atoms.  
The overall agreement with existing literature data is excellent for the entire region, with X-only results 
being very close to HF. An analysis of energy ordering is offered in terms of traditional correlation 
diagram, and individual energy components. 

To the best of our knowledge, this is the first reporting of excited state of confined atoms in very strong 
confinement ($r_c \leq 0.5$) region, excepting the work of \cite{aquino06}, which published results for ground and 
singly excited (1s2s $^3$S, up to $r_c= 2$) states of He. The correlation contribution to energy in smaller 
box ($r_c \leq 1$) is rather less dramatic than in a free atom. Thus, it is possible to obtain quite 
accurate results for a given state, provided the exchange contribution is properly accounted
for, which, of course, lies behind the general success of this approach. The two correlation functionals
(Wigner and LYP) behave quite differently as box size is changed. In the former, its magnitude gradually increases 
with confinement strength, whereas LYP shows a reduction until the appearance of a minimum at a certain
$r_c$, followed by a steep rise. In free limit, LYP appears to perform better than Wigner. In the opposite limit, 
however, Wigner seems to have an edge over LYP, reflecting a trend which is qualitatively similar to that 
observed in literature. Confinement induces interesting energy level crossings, in conformity with Hund's rule.
When pressure enhances, the asymptotic behavior of states are greatly affected; energy level goes from an 
atomic mean field theory ($r_c \rightarrow \infty$) to particle in a hard-sphere model ($r_c \rightarrow 0$). 
These crossings are mainly attributed to X-only energy. For more accurate results, better correlation energy 
functionals need to be designed and employed; one such functional reported in \cite{vyboishchikov17} is 
currently being pursued by us. Further application of the method is being made to other atoms as 
well as more realistic confinement scenario (such as \emph{soft or penetrable} potentials). The results on 
these works are as encouraging as reported here. However to limit the size of this article, it appears prudent 
to communicate them separately in future. It would also be worthwhile to further probe the critical radii 
($r_c$ where energies become zero), influence of 
electric and magnetic field through dynamical study, information theoretical analysis, etc. 

\section{Acknowledgement}
SM is grateful to IISER Kolkata for a Senior Research Fellowship. AKR gratefully acknowledges BRNS, Mumbai, 
India (sanction order: 58/14/03/2019-BRNS/10255) for financial support. We thank Dr. Neetik
Mukherjee for giving a critical reading of the manuscript.  

\bibliography{ref_s1}
\end {document}